\documentclass[12pt,reqno]{amsart}
\usepackage{graphicx}
\usepackage{amssymb,amsmath}
\usepackage{amsthm}
\usepackage{color,graphicx}
\usepackage{hyperref}
\usepackage{color}
\usepackage{epstopdf}
\usepackage{caption}
\usepackage{subcaption}
\usepackage[dvips]{epsfig}

\newcommand{\cn}{{\rm \,cn}}
\newcommand{\sn}{{\rm \,sn}}

\newcommand{\R}{{\mathbb R}}

\begin{document}
	\title[Periodic waves in the modified Korteweg--de Vries equation]{Periodic waves of the modified KdV equation as minimizers of a new variational problem}
	
	\author{Uyen Le}
	\address[U. Le]{Department of Mathematics and Statistics, McMaster University,
		Hamilton, Ontario, Canada, L8S 4K1}
	\email{leu@mcmaster.ca}
	
	\author{Dmitry E. Pelinovsky}
	\address[D. Pelinovsky]{Department of Mathematics and Statistics, McMaster University,
		Hamilton, Ontario, Canada, L8S 4K1 }
	\email{dmpeli@math.mcmaster.ca}
	
	
	
	\maketitle
	
	\begin{abstract}
		Periodic waves of the modified Korteweg-de Vries (mKdV) equation are identified in the context of a new variational problem with two constraints. The advantage of this variational problem is that its  non-degenerate local minimizers are stable in the time evolution of the mKdV equation, whereas the saddle points are unstable. We explore the analytical representation of periodic waves given by Jacobi elliptic functions and compute numerically critical points of  the constrained variational problem. A broken pitchfork bifurcation of three smooth solution families is found. Two families represent (stable) minimizers of the constrained variational problem and one family represents (unstable) saddle points.
	\end{abstract}
	
\bigskip

	
\section{Introduction}
	
We address travelling periodic waves of the modified Korteweg--de Vries (mKdV) equation which we take in the normalized form
\begin{equation}\label{mKdv}
		u_t+ 6u^2u_x + u_{xxx} =0.
\end{equation}
For the sake of clarity, we normalize the wave period to $2\pi$ 
and denote the Sobolev spaces of $2\pi$-periodic functions by $H^k_{\rm per}$ for $k \in \mathbb{N}$ with $H^0_{\rm per} \equiv L^2_{\rm per}$ for $k = 0$.
 
Travelling waves of the form $u(t,x)= \psi(x-ct)$ satisfy the stationary equation
\begin{equation}\label{stationary}
		-\psi''+ c\psi+ b= 2\psi^3,
\end{equation}
	where $\psi(x): [0, 2\pi]\mapsto \R$ is the wave profile, $c$ is the wave speed, and $b$ is the constant of integration. Travelling periodic waves of the mKdV equation (\ref{mKdv}) play a fundamental role in physical applications, e.g. for dynamics of the internal waves in seas and oceans 
	\cite{Grimshaw,Pelinovsky,Sutherland}. Period function for periodic solutions of the stationary 
	equation (\ref{stationary}) has been studied in \cite{Chow,Gavrilov,Yaga}. Stability of periodic waves in the defocusing version 
	of the mKdV equation has been studied in all details \cite{Niv}. We consider the focusing version of the mKdV equation (\ref{mKdv}).
	
There exist two families of periodic solutions to the stationary equation (\ref{stationary}) for $b = 0$:  dnoidal waves with sign-definite profile $\psi$ and cnoidal waves with sign-indefinite profile $\psi$. Spectral and orbital stability of these periodic waves has been explored in the recent literature \cite{angulo1,AN1,DK}. While sign-definite dnoidal waves are stable for all speeds, sign-indefinite cnoidal waves are stable for smaller speeds $c$ and unstable for larger speeds $c$ \cite{DK}. 

Compared to these definite results, the stationary mKdV equation (\ref{stationary}) with $b \neq 0$ has more general families of periodic waves expressed as two rational functions of Jacobi elliptic functions \cite{CPgardner}. Stability of a particular family of positive periodic waves with $b\neq0$ has been proven in \cite{AP2}, but no general results on stability of these periodic waves are available in the literature 
to the best of our knowledge. A new variational formulation 
of the periodic waves with $b \neq 0$ was developed in our previous works 
with F. Natali \cite{NLP,NLP2} (see also the follow-up work \cite{Natali}).

The purpose of this paper is to explore the new variational formulation 
of travelling periodic waves 
and to detect numerically which periodic waves with $b \neq 0$ are stable and 
which are unstable in the time evolution of the mKdV equation (\ref{mKdv}).
	
Since the mKdV equation (\ref{mKdv}) admits the following 
conserved quantities on the $2\pi$-periodic domain:
\begin{equation}
		E(u)= \frac{1}{2} \oint \left[(u')^2 - u^4\right]dx,\quad
		F(u)= \frac{1}{2} \oint u^2dx,\quad
		M(u)= \oint u dx,
		\label{cons-quant}
	\end{equation}
the stationary equation \eqref{stationary} is the Euler--Lagrange equation for the action functional
\begin{equation}
\label{action functional}
		G_{c,b}(u) := E(u) + cF(u) +bM(u). 
\end{equation} 
We refer to $E(u)$, $F(u)$, and $M(u)$ as the energy, momentum, and mass, 
respectively.

The standard variational formulation for stability of periodic waves is to find minimizers of energy $E(u)$ in $H^1_{\rm per}$ subject to the fixed momentum $F(u)$ and mass $M(u)$ \cite{bronski,haragus,johnson,pelinovsky}.
Parameters $c$ and $b$ of the stationary equation (\ref{stationary}) are Lagrange multipliers of the action (\ref{action functional}). Unfortunately, this formulation may suffer from non-smooth dependence of the minimizers from Lagrange multipliers $(c,b)$ as discussed in \cite{NLP,NLP2} after \cite{hur}. This breakdown of the variational theory happens at the bifurcation points for which the Hessian operator for $G_{c,b}(u)$ admits a zero eigenvalue, where $\mathcal{L} = G_{c,b}''(\psi) : H^2_{\rm per}\subset L^2_{\rm per} \mapsto L^2_{\rm per}$ is given by 
\begin{equation}\label{operator}
\mathcal{L} = -\partial^2_x + c -6\psi^2.
\end{equation}
In \cite{NLP2} (based on the previous work \cite{NLP} in the case of quadratic nonlinearities), we have proposed a new variational approach to characterize 
the periodic waves of the stationary equation (\ref{stationary}) as minimizers of the following constrained variational problem:
\begin{equation}\label{2constraints}
		r_{c,m}:= \inf_{u\in H^{1}_{\rm per}} \left\lbrace \mathcal{B}_c(u): \quad \oint u^4 dx=1, \quad \frac{1}{2\pi} \oint  u dx = m \right\rbrace,
\end{equation}
where 
	\begin{equation}\label{functionalB}
	\mathcal{B}_c(u):=\frac{1}{2}\oint  \left[(u')^2 +cu^2\right] dx 
	\end{equation}
It was shown in \cite[Appendix B]{NLP2} that the minimizer exists for every $c\in (-1, \infty)$ and every $m\in [-m_0, m_0]$, where $m_0 := (2\pi)^{-1/4}$. The minimizer has one maximum and one minimum on the $2\pi$-periodic domain if $m\in (-m_0, m_0)$ and is given by the constant solution if $m = \pm m_0$. The minimizer $\chi \in H^1_{\rm per}$ such that $r_{c,m} = \mathcal{B}_c(\chi)$ gives the solution of the stationary equation (\ref{stationary}) by using the scaling transformation 
\begin{equation}
\label{multipliers-direct}
\psi = \chi \frac{\sqrt{\mathcal{B}_c(\chi) - \pi c m^2}}{\sqrt{1 - m \oint \chi^3 dx}},
\end{equation}
and it was shown in \cite{NLP2} that $\mathcal{B}_c(\chi) - \pi c m^2 > 0$ 
and $1 - m \oint \chi^3 dx > 0$. The inverse transformation is given by 
$\chi = \psi/\| \psi \|_{L^4}$, hence 
\begin{equation}
\label{multipliers}
m = \frac{1}{2\pi \| \psi \|_{L^4}} \oint \psi dx.
\end{equation}

The family of sign-indefinite cnoidal waves for the stationary equation (\ref{stationary}) with $b = 0$ was recovered in \cite{NLP2} from the variational problem (\ref{2constraints}) for $c \in (-1,\infty)$ and $m = 0$ in the space of odd periodic functions. It was found that the family is smooth with respect to parameter $c$ 
but there exists a bifurcation point $c_0 \approx 1.425$ such that the sign-indefinite wave is not a minimizer 
of the variational problem (\ref{2constraints}) for $c \in (c_0,\infty)$ and $m = 0$. The bifurcation point $c_0$ coincides with the stability threshold 
found in \cite{DK}. 

Similar study in the case of quadratic nonlinearity in \cite{NLP} also showed 
that the family of minimizers of the new variational problem for the travelling periodic waves with zero mean remains smooth with respect to the wave speed $c$.

Another example when stability of periodic waves have to be studied outside the standard variational theory was investigated in \cite{Geyer} within the framework of the Camassa--Holm equation. An alternative Hamiltonian structure was used in order to provide smooth continuation of the periodic waves and the stability conclusion. \\

Let us now explain the organization of this paper.
 
In Section 2, we develop the stability theory for the travelling periodic waves given by non-degenerate local minimizers and saddle points of the variational problem (\ref{2constraints}). 
We derive a precise stability criterion for a non-degenerate local minimizer 
and a precise instability criterion for a saddle point of the variational problem (\ref{2constraints}).

In Section 3, we perform the numerical search of critical points of the variational problem (\ref{2constraints}). We show that the global minimizers remain smooth in $(c,m)$ for every $c \in (-1,\infty)$ and $m \in (0,m_0)$. Besides the smooth family of global minimizers, there exist two other families of periodic waves in a subset of the region $c \in (-1,\infty)$ and $m \in (0,m_0)$: one family contains local minimizers and the other family contains saddle points of the variational problem (\ref{2constraints}). The two families disappear at the fold bifurcation point $c_*(m)$. 
When $m \to 0$, $c_*(m) \to c_0 \approx 1.425$, where the three families are connected in the pitchfork bifurcation observed in \cite{NLP2}. 
No other solution families have been identified in the numerical search.

Computing the stability criterion numerically, we show that the two families of minimizers are stable in the time evolution of the mKdV equation (\ref{mKdv}) whereas the only family of saddle points is unstable. These results generalize the result of 
\cite{NLP2} obtained for $m = 0$.

Section 4 concludes the paper with a summary and a discussion of further questions. 

\section{Stability theory for non-degenerate critical points}

Let $\chi \in H^1_{\rm per}$ be a minimizer of the variational problem (\ref{2constraints}) for $c \in (-1,\infty)$ and $m \in (0,m_0)$
which always exists by Theorem 2.1 and Proposition 6.5 in \cite{NLP2}. 
Let $\psi \in H^1_{\rm per}$ be obtained by 
means of the transformation (\ref{multipliers-direct}). 
Then, $\psi$ satisfies the 
stationary equation (\ref{stationary}) with uniquely defined function 
\begin{equation}
\label{b-function}
b = b(c,m) := \frac{1}{\pi} \oint \psi^3 dx - c m \| \psi \|_{L^4}.
\end{equation}
The profile $\psi$ has exactly one maximum and one minimum point on the $2\pi$-periodic domain. The main assumption on the minimizer $\chi$ is given as follows.

\vspace{0.2cm}
\centerline{\fbox{\parbox[cs]{0.95\textwidth}{
Assume that $\chi$ is a non-degenerate minimizer of the variational problem (\ref{2constraints}) module 
to the translational symmetry: $\chi(x) \mapsto \chi(x+x_0)$ for every $x_0 \in \mathbb{R}$.
}}}
\vspace{0.2cm}

For the solution $\psi$ of the stationary equation (\ref{stationary}) given by (\ref{multipliers-direct}), this assumption implies that the Hessian operator $\mathcal{L}$ in (\ref{operator}) restricted to the orthogonal complement of $\{ 1, \psi^3\}$ in $L^2_{\rm per}$ 
is positive and admits a simple zero eigenvalue with the eigenfunction $\partial_x \psi$. For the sake of notations, we denote the restriction of 
$\mathcal{L}$ to $\{ 1, \psi^3 \}^{\perp}$ in $L^2_{\rm per}$ by $\mathcal{L}|_{\{1,\psi^3\}^{\perp}}$. With the standard notations of $n(\mathcal{L})$ and $z(\mathcal{L})$ for the number of negative eigenvalues  and the multiplicity of the zero eigenvalue of a self-adjoint operator $\mathcal{L}$,  we express the main assumption in the form:
\begin{equation}
\label{assumption}
n(\mathcal{L}|_{\{1,\psi^3\}^{\perp}}) = 0, \quad z(\mathcal{L}|_{\{1,\psi^3\}^{\perp}}) = 1.
\end{equation}
By using the implicit function argument (similar to Lemma 2.8 in \cite{NLP2}), it is easy to prove that the assumption (\ref{assumption}) implies smoothness of 
the family of minimizers $\chi$ in $(c,m)$. It follows then with the help of (\ref{multipliers-direct}) and (\ref{b-function}) that the function $b$ is smooth in $(c,m)$. Hence, it can be differentiated in $(c,m)$, from which we can characterize the range of $\mathcal{L}$ by using 
\begin{eqnarray}
\label{range-1} && \mathcal{L}1 = c -6\psi^2,\\
\label{range-2} && \mathcal{L}\psi = -b -4\psi^3,\\
\label{range-3} && \mathcal{L}\partial_c\psi = -\partial_c b -\psi, \label{L inverse psi}\\
\label{range-4} && \mathcal{L}\partial_m\psi = -\partial_m b.
\end{eqnarray}
We recall from \cite{hur} (see also Proposition 2.5 in \cite{NLP2}) 
that  ${\rm Ker}(\mathcal{L}) = {\rm span}(\partial_x \psi)$ if and only if 
$\lbrace 1, \psi, \psi^2 \rbrace \in \textrm{Range}(\mathcal{L})$. 
By using this result, it follows from (\ref{range-1}), (\ref{range-3}), and (\ref{range-4}) that $z(\mathcal{L}) = 1$ if and only if $\partial_m b \neq 0$.

We also recall that since $\partial_x \psi$ has exactly two zeros on the $2\pi$-periodic domain, Sturm's nodal theory from \cite{hur} (see also Proposition 2.4 in \cite{NLP2}) implies that $\mathcal{L}$ is not positive and admits at least one negative eigenvalue: $n(\mathcal{L}) \geq 1$. 
Since $\psi$ is related to a minimizer of the variational problem (\ref{2constraints}) with two constraints: $n(\mathcal{L}) \leq 2$. Hence, we 
have $1 \leq n(\mathcal{L}) \leq 2$.

Next we count $n(\mathcal{L})$ based on the standard count of eigenvalues of a self-adjoint operator under two orthogonal constraints (see Lemma 2.13 in \cite{NLP2} and Theorem 4.1 in \cite{Pel-book}). We compute the limit $\lambda \to 0$ of the following matrix:
\begin{equation}
P(\lambda) := \left[\begin{array}{cc}\langle (\mathcal{L} - \lambda I)^{-1} \psi^3,\psi^3 \rangle &
\langle (\mathcal{L} - \lambda I)^{-1}\psi^3, 1 \rangle \\
\langle (\mathcal{L} - \lambda I)^{-1} 1, \psi^3 \rangle &
\langle (\mathcal{L} - \lambda I)^{-1}1,1\rangle
\end{array}\right], \quad \lambda \notin \sigma(\mathcal{L}).
\end{equation}
If $\partial_m b \neq 0$, it follows from (\ref{range-2}) and (\ref{range-4}) that 
\begin{align*}
& \langle \mathcal{L}^{-1}1,1\rangle = -\frac{1}{\partial_mb}\partial_m\left(\oint \psi dx\right),\\
& \langle \mathcal{L}^{-1} 1, \psi^3 \rangle = -\frac{1}{4\partial_mb}\partial_m\left(\oint \psi^4 dx\right),\\
& \langle \mathcal{L}^{-1}\psi^3, 1 \rangle = - \frac{1}{4}\oint \psi dx- \frac{b}{4}\langle \mathcal{L}^{-1}1,1\rangle,\\
& \langle\mathcal{L}^{-1} \psi^3,\psi^3 \rangle = -\frac{1}{4}\oint\psi^4dx- \frac{b}{4} \langle \mathcal{L}^{-1}1,\psi^3 \rangle,	
\end{align*}
which yields
\begin{align*}
\lim_{\lambda \to 0} \det(P(\lambda)) &=  \langle \mathcal{L}^{-1} \psi^3,\psi^3 \rangle \; \langle \mathcal{L}^{-1}1,1\rangle - \langle \mathcal{L}^{-1} \psi^3,1\rangle \; \langle \mathcal{L}^{-1} 1, \psi^3 \rangle,\\
&= \frac{1}{4\partial_mb}\left[ \left(\oint\psi^4 dx\right)\partial_m\left(\oint \psi dx\right) - \frac{1}{4}\left(\oint \psi dx\right)\partial_m\left(\oint \psi^4 dx\right)  \right],\\
&= \frac{1}{4\partial_m b}\left(\oint \psi^4dx\right)^{\frac{5}{4}} \partial_m\left(\frac{\oint \psi dx}{\left(\oint \psi^4dx\right)^{\frac{1}{4}}}\right),\\
&= \frac{\pi}{2\partial_m b}\| \psi\|_{L^4}^5,
\end{align*}
where the relation (\ref{multipliers}) has been used.
Thus, we conclude that the sign of $\lim\limits_{\lambda \to 0} \det(P(\lambda))$ coincides with 
the sign of $\partial_m b$. 

Recall the counting formulas (see  Proposition 2.12 in \cite{NLP2}):
\begin{equation}
\left\{ \begin{array}{l} 
n(\mathcal{L}|_{\{1,\psi^3\}^{\perp}}) = n(\mathcal{L}) - n_0 - z_0 = 0, \\
z(\mathcal{L}|_{\{1,\psi^3\}^{\perp}}) = z(\mathcal{L}) + z_0 - z_{\infty} = 1,
\end{array} \right.
\label{count}
\end{equation}
where $z_0$ is multiplicity of zero eigenvalue of $\lim\limits_{\lambda \to 0} P(\lambda)$, $n_0$ is the number of negative eigenvalues of $\lim\limits_{\lambda \to 0} P(\lambda)$, and 
$z_{\infty}$ is the number of eigenvalues $P(\lambda)$ diverging to infinity as $\lambda \to 0$.

\begin{itemize}
	\item If $\partial_m b < 0$, then $\lim\limits_{\lambda \to 0} P(\lambda)$ has one negative eigenvalue so that $n_0 = 1$, $z_0 = z_{\infty} = 0$ implying from (\ref{count}) that $n(\mathcal{L}) = 1$ and $z(\mathcal{L}) = 1$.

	\item If $\partial_m b > 0$, then $\lim\limits_{\lambda \to 0} P(\lambda)$ is either  positive or negative, but the former case would imply from (\ref{count}) that $\mathcal{L}$ is positive, in contradiction with $n(\mathcal{L}) \geq 1$. Hence $\lim\limits_{\lambda \to 0} P(\lambda)$ is negative with $n_0 = 2$, $z_0 = z_{\infty} = 0$ implying from (\ref{count}) that $n(\mathcal{L}) = 2$ and $z(\mathcal{L}) = 1$.

	\item If $\partial_m b = 0$, then $z_{\infty} = 1$, $z_0 = 0$ implying from 
	(\ref{count}) that $z(\mathcal{L}) = 2$. Thus, one of the two negative eigenvalues of $\mathcal{L}$ for $\partial_m b > 0$ passes zero at $\partial_m b = 0$ and becomes a positive eigenvalue of $\mathcal{L}$ for $\partial_m b < 0$.
\end{itemize}

These computations are summarized as follows:
\begin{equation}
n(\mathcal{L}) = \left\{ \begin{array}{ll} 2, \quad & \mbox{\rm if } \partial_m b > 0, \\
1, \quad & \mbox{\rm if } \partial_m b \leq 0, \end{array}\right. \qquad 
z(\mathcal{L}) = \left\{ \begin{array}{ll} 2, \quad & \mbox{\rm if } \partial_m b = 0, \\
1, \quad & \mbox{\rm if } \partial_m b \neq 0. \end{array}\right.
\label{Morse}
\end{equation}

Next we determine the stability of minimizers in the time evolution 
of the mKdV equation (\ref{mKdv}) by computing 
$n(\mathcal{L}|_{\{1,\psi\}^{\perp}})$ and $z(\mathcal{L}|_{\{1,\psi\}^{\perp}})$ 
and by using the stability criteria  from \cite{haragus}:
\begin{itemize}
	\item The periodic wave with profile $\psi$ is stable if 
	\begin{equation}
	n(\mathcal{L}|_{\{1,\psi\}^{\perp}}) = 0 \qquad \mbox{\rm and} \qquad  z(\mathcal{L}|_{\{1,\psi\}^{\perp}}) = 1
	\label{stability}
	\end{equation}
	\item The periodic wave with profile $\psi$ is unstable if 
	\begin{equation}
	n(\mathcal{L}|_{\{1,\psi\}^{\perp}}) = 1 \qquad \mbox{\rm and} \qquad  z(\mathcal{L}|_{\{1,\psi\}^{\perp}}) = 1.
	\label{instability}
	\end{equation}
\end{itemize}
Similar to Theorem 2.14 in \cite{NLP2}, we compute the limit $\lambda \to 0$ of the following matrix:
\begin{equation}
D(\lambda) := \left[\begin{array}{cc}\langle (\mathcal{L} - \lambda I)^{-1} \psi,\psi \rangle &
\langle (\mathcal{L} - \lambda I)^{-1}\psi, 1 \rangle \\
\langle (\mathcal{L} - \lambda I)^{-1} 1, \psi \rangle &
\langle (\mathcal{L} - \lambda I)^{-1}1,1\rangle
\end{array}\right], \quad \lambda \notin \sigma(\mathcal{L}).
\end{equation}
If $\partial_m b \neq 0$, it follows from (\ref{range-3}) and (\ref{range-4}) that 
\begin{align*}
\langle \mathcal{L}^{-1}1,1\rangle &= -\frac{1}{\partial_mb}\partial_m\left(\oint \psi dx\right),\\
\langle \mathcal{L}^{-1} 1, \psi \rangle&= -\frac{1}{2\partial_mb}\partial_m\left(\oint \psi^2 dx\right),\\
\langle \mathcal{L}^{-1}\psi, 1 \rangle &= - \partial_c\left(\oint \psi dx\right)- \partial_cb \;\langle \mathcal{L}^{-1}1,1\rangle,\\
\langle\mathcal{L}^{-1} \psi,\psi \rangle &= -\frac{1}{2}\partial_c\left(\oint\psi^2dx\right)- \partial_cb \; \langle \mathcal{L}^{-1}1,\psi \rangle,		
\end{align*}
which yields
\begin{align*}
\lim_{\lambda \to 0} \det(D(\lambda)) &=  \langle \mathcal{L}^{-1} \psi,\psi \rangle \; \langle \mathcal{L}^{-1}1,1\rangle - \langle \mathcal{L}^{-1} \psi,1\rangle \; \langle \mathcal{L}^{-1} 1, \psi \rangle,\\
&= \frac{1}{2\partial_m b} \left[ \partial_c\left(\oint \psi^2 dx\right) \partial_m \left(\oint\psi dx\right) - \partial_m\left(\oint \psi^2dx\right)\partial_c\left(\oint \psi dx\right)\right] \\
&= \frac{1}{2\partial_m b}\begin{vmatrix}
\partial_c \mathcal{F}(c,m) & \partial_m \mathcal{F}(c,m)  \\ 
\partial_c \mathcal{M}(c,m) & \partial_m \mathcal{M}(c,m) 
\end{vmatrix},
\end{align*}
where $\mathcal{F}(c,m) := F(\psi)$ and $\mathcal{M}(c,m) := M(\psi)$ are computed from the two conserved quantities in (\ref{cons-quant}) at the family of periodic waves with the profile $\psi$ that depends on parameters $(c,m)$. 
Note that the determinant in the last expression is the Jacobian
of the transformation $(c,m) \mapsto (\mathcal{F},\mathcal{M})$. 

We shall now use the counting formulas:
\begin{equation}
\left\{ \begin{array}{l} 
n(\mathcal{L}|_{\{1,\psi\}^{\perp}}) = n(\mathcal{L}) - n_0 - z_0, \\
z(\mathcal{L}|_{\{1,\psi\}^{\perp}}) = z(\mathcal{L}) + z_0 - z_{\infty},
\end{array} \right.
\label{count-stability}
\end{equation}
where $z_0$, $n_0$, and $z_{\infty}$ have the same meaning as in (\ref{count}) 
but for the matrix $D(\lambda)$.

\begin{itemize}
	\item If $\partial_m b < 0$, then it follows from (\ref{Morse}) that $n(\mathcal{L}) = 1$ and $z(\mathcal{L}) = 1$ so that the stability criterion (\ref{stability}) is satisfied if and only if $n_0 = 1$, $z_0 = z_{\infty} = 0$, which is true if and only if the Jacobian of the transformation $(c,m) \mapsto (\mathcal{F},\mathcal{M})$ is strictly positive. 

	\item If $\partial_m b > 0$, then it follows from (\ref{Morse}) 
	that $n(\mathcal{L}) = 2$ and $z(\mathcal{L}) = 1$ so that the stability criterion (\ref{stability}) is satisfied  if and only if 
	$n_0 = 2$, $z_0 = z_{\infty} = 0$, that is, $\langle \mathcal{L}^{-1}1,1\rangle < 0$ and the Jacobian of the transformation $(c,m) \mapsto (\mathcal{F},\mathcal{M})$ is strictly positive. Since $\langle \mathcal{L}^{-1}1,1\rangle$ is the same diagonal term of both $\lim\limits_{\lambda \to 0} P(\lambda)$ and $\lim\limits_{\lambda \to 0} D(\lambda)$ whereas the former is strictly negative, the first condition of $\langle \mathcal{L}^{-1}1,1\rangle < 0$ is satisfied.

	\item If $\partial_m b = 0$, then it follows from (\ref{Morse})  that $n(\mathcal{L}) = 1$ and $z(\mathcal{L}) = 2$ so that $\det D(\lambda)$ is singular in the limit $\lambda \to 0$. Hence $z_{\infty} = 1$ and one of the two negative eigenvalues of $\lim\limits_{\lambda \to 0} D(\lambda)$ for $\partial_m b > 0$ diverges to infinity as $\partial_m b \to 0$, whereas the other eigenvalue remains negative if and only if the Jacobian of the transformation $(c,m) \mapsto (\mathcal{F},\mathcal{M})$ is strictly positive. This yields $n_0 = 1$, $z_0 = 0$, and hence the stability criterion (\ref{stability}).
\end{itemize}

The stability criterion in all three cases can be summarized as follows. 

\vspace{0.2cm}
\centerline{\fbox{\parbox[cs]{0.95\textwidth}{
			Let $\psi \in H^1_{\rm per}$ be a solution 
			of the stationary equation (\ref{stationary}) satisfying (\ref{assumption}).
			The periodic wave with the profile $\psi$ is stable 
			in the time evolution of the mKdV equation (\ref{mKdv}) if 
\begin{equation}\label{stability criterion}
\begin{vmatrix}
\partial_c \mathcal{F}(c,m) & \partial_m \mathcal{F}(c,m)  \\ 
\partial_c \mathcal{M}(c,m) & \partial_m \mathcal{M}(c,m) 
\end{vmatrix} >0.
\end{equation}
}}}
\vspace{0.2cm}

Note that the assumption (\ref{assumption}) is satisfied for the solution $\psi \in H^1_{\rm per}$ related to both the global and local non-degenerate minimizers $\chi \in H^1_{\rm per}$ of the variational problem (\ref{2constraints}). 
Let us now consider the non-degenerate saddle points. 

\vspace{0.2cm}
\centerline{\fbox{\parbox[cs]{0.95\textwidth}{
			Assume that $\chi$ is a non-degenerate saddle point of the variational problem (\ref{2constraints}) module to the translational symmetry: $\chi(x) \mapsto \chi(x+x_0)$ for every $x_0 \in \mathbb{R}$ with exactly one negative direction in $H^1_{\rm per}$ under the two constraints. 
}}}
\vspace{0.2cm}			
			
The main assumption for the corresponding solution $\psi \in H^1_{\rm per}$ of the stationary equation (\ref{stationary}) can be expressed in the form:
\begin{equation}
\label{assumption-saddle}
n(\mathcal{L}|_{\{1,\psi^3\}^{\perp}}) = 1, \quad z(\mathcal{L}|_{\{1,\psi^3\}^{\perp}}) = 1.
\end{equation}

The non-degeneracy of saddle point implies smoothness of the function $b$ in $(c,m)$. As a result, the same count of $n(\mathcal{L})$ can be performed based on $\lim\limits P(\lambda)$ with exactly the same expression for $\lim\limits_{\lambda \to 0} \det(P(\lambda))$ if $\partial_m b = 0$.
Since $\lim\limits_{\lambda \to 0} \det(P(\lambda)) \neq 0$, it follows from (\ref{count}) with $z_0 = z_{\infty} = 0$ that $n(\mathcal{L}) = 2$ if $\partial_m b < 0$ for which $n_0 = 1$. Since 
$1 \leq n(\mathcal{L}) \leq 2$ is no longer true for the saddle points, 
the case with $\partial_m b > 0$ may either give $n(\mathcal{L}) = 1$ or $n(\mathcal{L}) = 3$. To avoid ambiguity, we only consider the saddle points with $\partial_m b < 0$.

It follows from (\ref{count-stability}) that if $\partial_m b < 0$, 
then the instability criterion (\ref{instability}) is satisfied 
if and only if the Jacobian of the transformation $(c,m) \mapsto (\mathcal{F},\mathcal{M})$ is strictly positive for which $n_0 = 1$, $z_0 = z_{\infty} = 0$. The instability criterion can be summarized as follows. 

\vspace{0.2cm}
\centerline{\fbox{\parbox[cs]{0.95\textwidth}{
			Let $\psi \in H^1_{\rm per}$ be a solution 
			of the stationary equation (\ref{stationary}) satisfying (\ref{assumption-saddle}).
			The periodic wave with the profile $\psi$ is unstable 
			in the time evolution of the mKdV equation (\ref{mKdv}) if 
			$\partial_m b < 0$ and 
			\begin{equation}\label{instability criterion}
			\begin{vmatrix}
			\partial_c \mathcal{F}(c,m) & \partial_m \mathcal{F}(c,m)  \\ 
			\partial_c \mathcal{M}(c,m) & \partial_m \mathcal{M}(c,m) 
			\end{vmatrix} >0.
			\end{equation}
}}}
\vspace{0.2cm}

Although the stability and instability criteria (\ref{stability criterion}) and (\ref{instability criterion}) are only sufficient conditions, we show with the help of numerical approximations that these criteria cover all critical points of the variational problem (\ref{2constraints}).

\section{Numerical search of critical points of the variational problem}

To perform the numerical search, we use the analytical representation of solutions to the stationary equation (\ref{stationary}) in terms of the Jacobi elliptic functions. Such representations are known in the literature; we refer to \cite{CPgardner} for precise details. 

One family of exact solutions is given by 
\begin{equation}\label{dnform}
\psi(x) = u_4 + \frac{(u_1-u_4)(u_2-u_4)}{(u_2-u_4)+(u_1-u_2)\sn^2(\nu x; k)},
\end{equation}
where the turning points $u_1$, $u_2$, $u_3$, $u_4$ satisfy the constraint 
\begin{equation}
\label{dn-constraint}
u_1+u_2+u_3+u_4 =0
\end{equation}
and define parameters $c$ and $b$ of the stationary equation (\ref{stationary}) by 
\begin{align}\label{dn parameters}
\begin{cases}
c = - (u_1u_2+u_1u_3+u_1u_4+u_2u_3+u_2u_4+u_3u_4),\\
b = \frac{1}{2} ( u_1u_2u_3+ u_1u_2u_4 + u_1u_3u_4 +u_2u_3u_4).
\end{cases}
\end{align} 
Parameters $\nu$ and $k$ of the solution (\ref{dnform}) are expressed by the relations:
$$
\nu = \frac{1}{2} \sqrt{(u_1- u_3)(u_2-u_4)},\quad
k = \frac{\sqrt{(u_1-u_2)(u_3-u_4)}}{\sqrt{(u_1- u_3)(u_2-u_4)}}.
$$ 
When a given value of $c \in (-1,\infty)$ is substituted into the first equation in (\ref{dn parameters}) and the additional constraint (\ref{dn-constraint}) is used, the family of solutions (\ref{dnform}) has two arbitrary parameters among the four turning points, which we choose to be $u_1$ and $u_2$. These two parameters are defined from two additional constraints: the period of $\psi$ must be normalized to $2 \pi$ by $2 K(k) = 2 \pi \nu$, where $K(k)$ is the complete elliptic integral, and the mean value and the $L^4$ norm of the solution $\psi$ must be related to a given value of $m \in (-m_0,m_0)$ by (\ref{multipliers}). Newton's method is used to solve these two constraints.

Another family of exact solutions is given by
\begin{equation}\label{cnform}
		\psi(x)= u_1 + \frac{(u_2 - u_1)(1-\cn(\nu x; k))}{1 - \cn(\nu x;k) + \delta (1 + \cn(\nu x;k))},
\end{equation}
with the turning points $u_1$, $u_2$, 
$u_3 = \alpha + i \beta$, $u_4 = \alpha - i \beta$. The turning points satisfy the constraint 
\begin{equation}
\label{cn-constraint}
u_1+u_2+2 \alpha =0.
\end{equation}
and define parameters $c$ and $b$ of the stationary equation (\ref{stationary}) by 
\begin{align}\label{cn parameters}
\begin{cases}
c = - (u_1u_2+2\alpha (u_1 + u_2) + \alpha^2 + \beta^2),\\
b = \alpha u_1u_2 + \frac{1}{2} (u_1 + u_2) (\alpha^2 + \beta^2).
\end{cases}
\end{align} 
Parameters $\nu$, $k$, and $\delta$ of the solution (\ref{cnform}) are expressed by the relations
$$
\delta = \frac{\sqrt{(u_2-\alpha)^2+\beta^2}}{\sqrt{(u_1-\alpha)^2+\beta^2}},\quad
\nu= \sqrt[4]{\left[(u_1-\alpha)^2+\beta^2\right] \left[(u_2-\alpha)^2+\beta^2\right]},
$$
and
$$
k = \frac{1}{\sqrt{2}} \sqrt{1- \frac{(u_1-\alpha)(u_2-\alpha)+\beta^2}{\sqrt{(u_1-\alpha)^2+\beta^2} \sqrt{(u_2-\alpha)^2+\beta^2}}},
$$ 
Again we use (\ref{cn-constraint}) to define $\alpha$ and 
the first equation in (\ref{cn parameters}) to define $\beta$ from $c \in (-1,\infty)$. The remaining parameters $u_1$ and $u_2$ are computed from two additional constraints: the period of $\psi$ must be normalized to $2 \pi$ 
by $4 K(k) = 2 \pi \nu$ and the mean value and the $L^4$ norm of the solution $\psi$ must be related to a given value of $m \in (-m_0,m_0)$ by (\ref{multipliers}). Newton's method is used to solve these two constraints.

\begin{figure}[htb!]
	\centering
	\begin{subfigure}[b]{0.5\textwidth}
		\centering
		\includegraphics[width=1\textwidth]{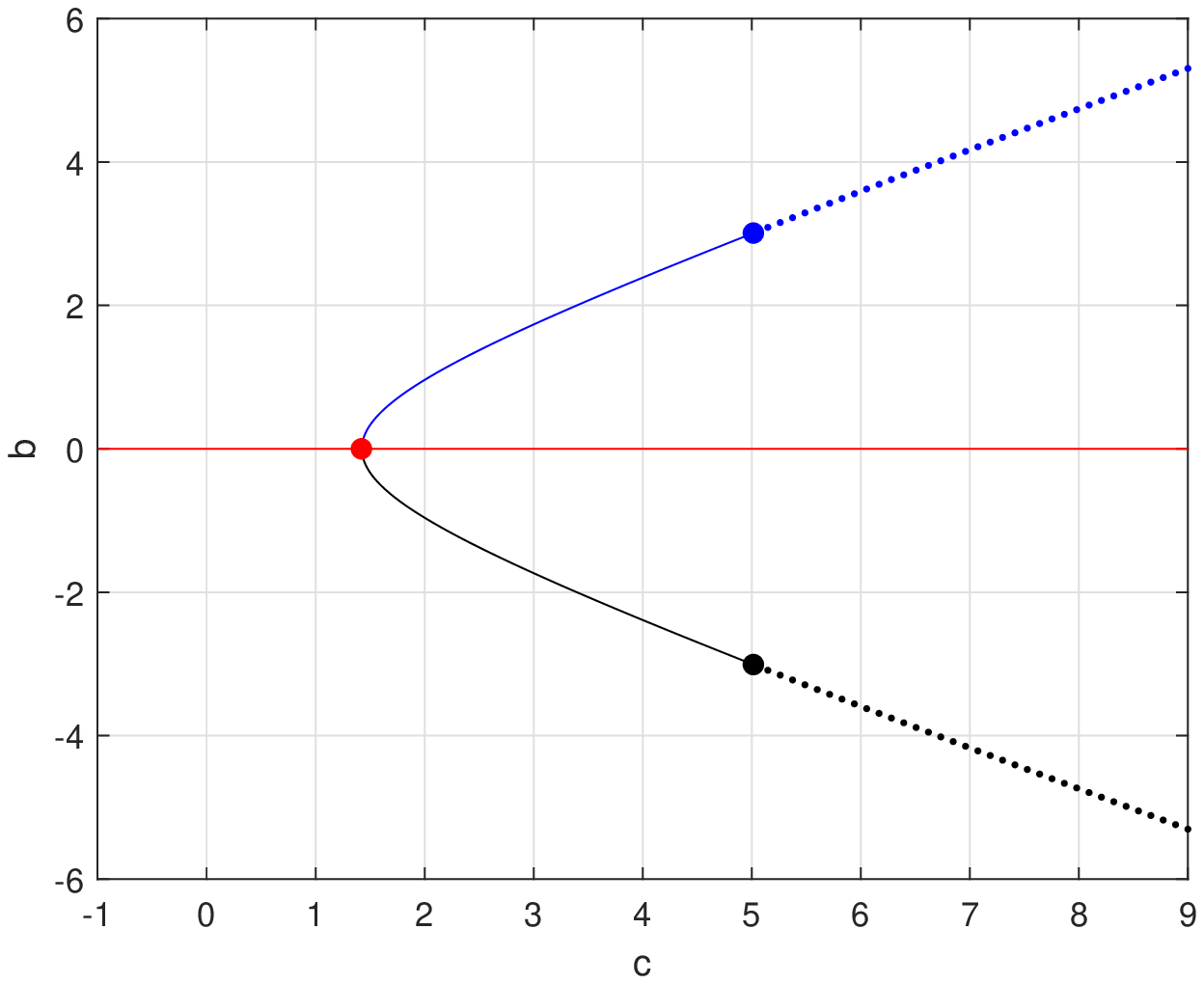}
		\caption{$m=0$}
		\label{fig:bifurA}
	\end{subfigure}
	\begin{subfigure}[b]{0.5\textwidth}
		\centering
		\includegraphics[width=1\textwidth]{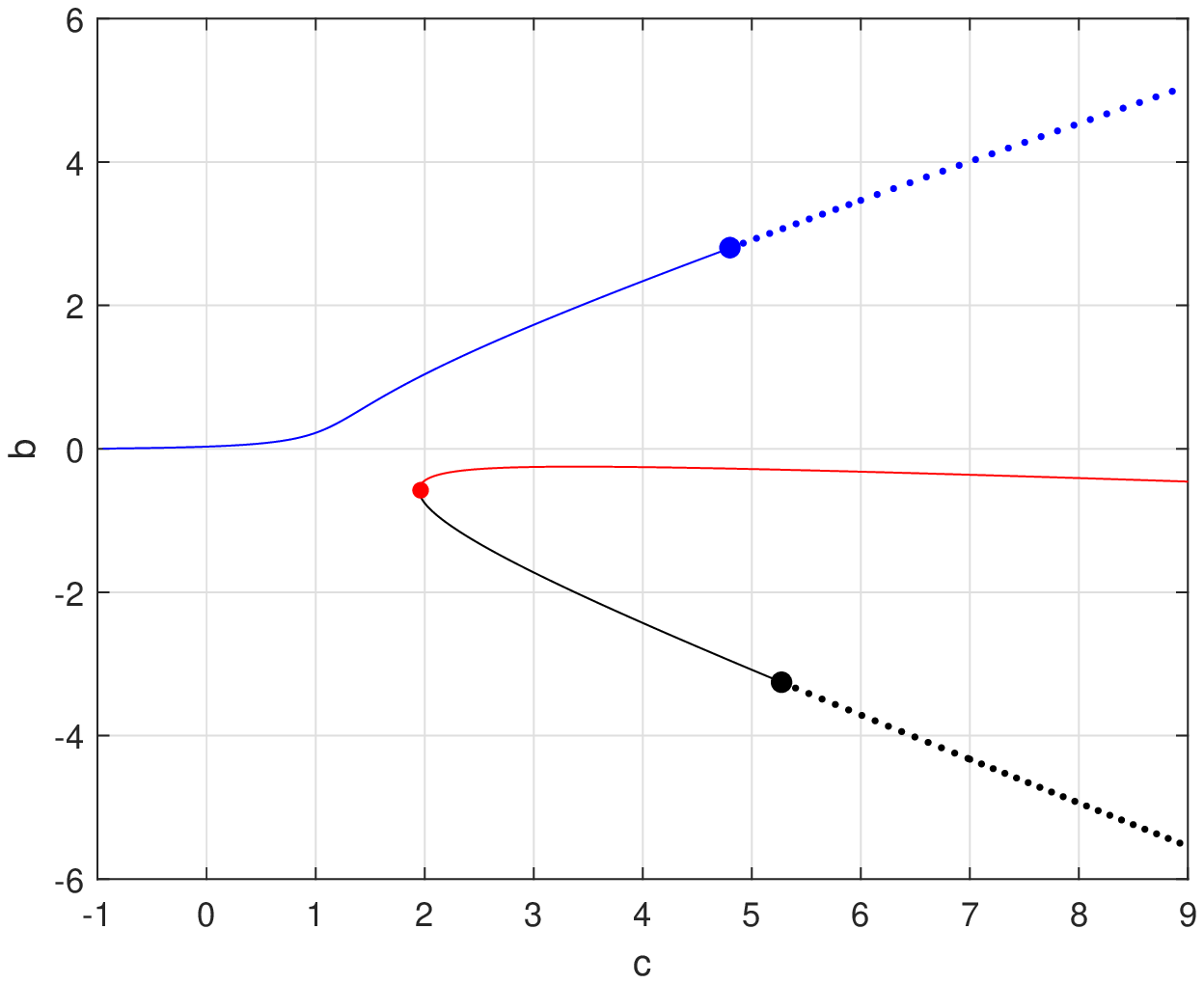}
		\caption{$m = 0.01$}
		\label{fig:bifurB}
	\end{subfigure}
	\begin{subfigure}[b]{0.5\textwidth}
		\centering
		\includegraphics[width=1\textwidth]{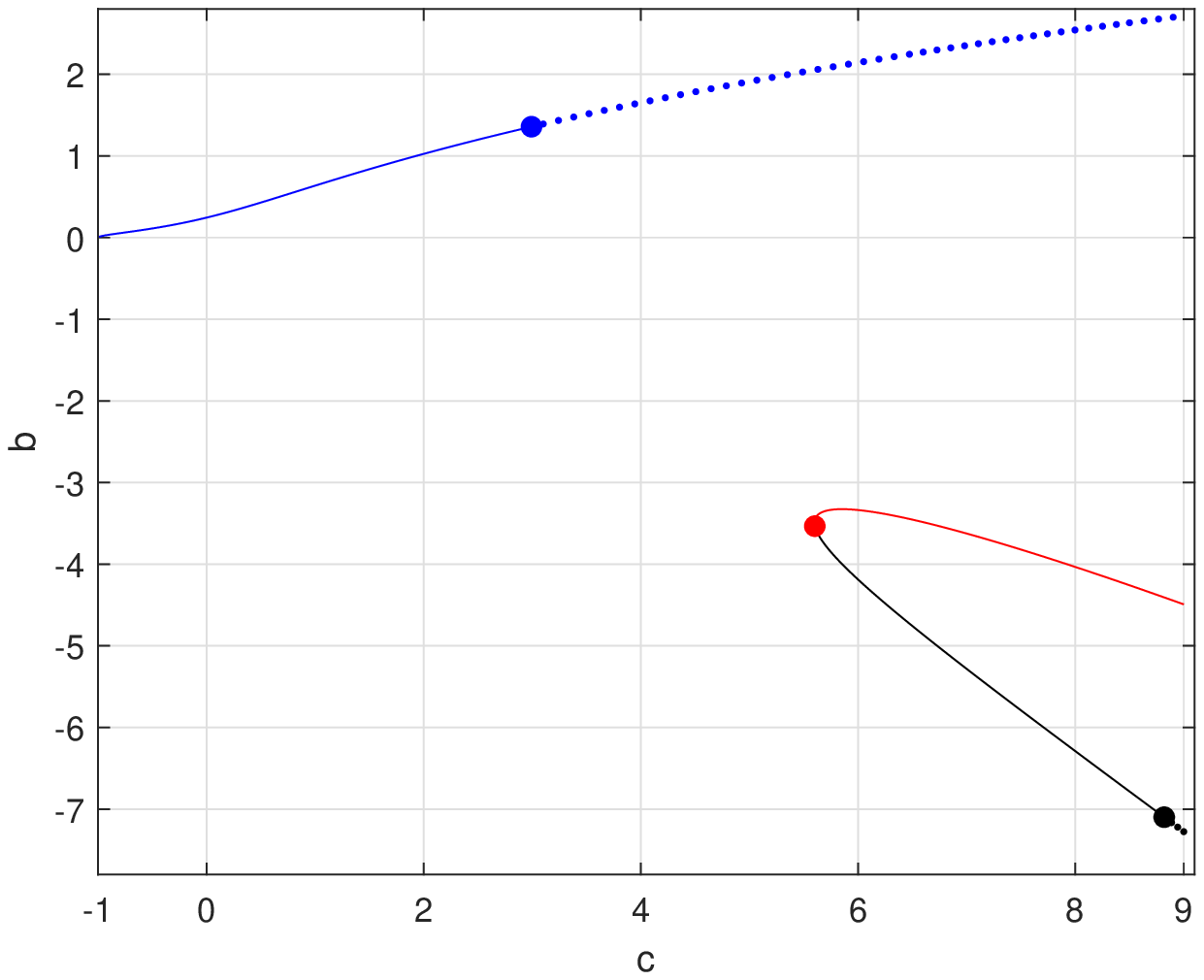}
		\caption{$m=0.1$}
		\label{fig:bifurC}
	\end{subfigure}
	\begin{subfigure}[b]{0.5\textwidth}
		\centering
		\includegraphics[width=1\textwidth]{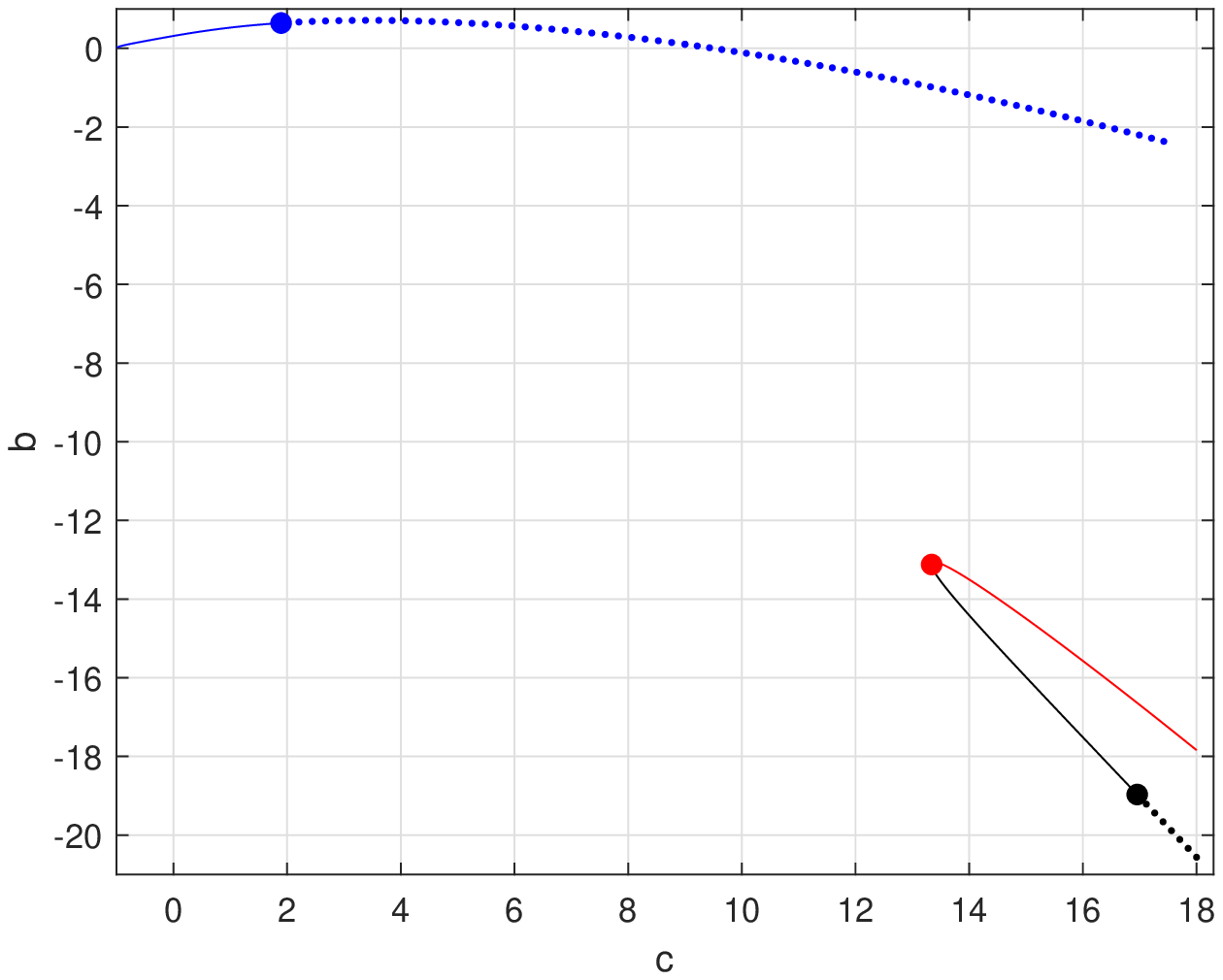}
		\caption{$m=0.2$}
		\label{fig:bifurD}
	\end{subfigure}
	\caption{Three solution families on the $(b,c)$ diagram for fixed values of $m$.}
	\label{fig:bifurs}
\end{figure}
	
Figure \ref{fig:bifurs} presents the main result in obtaining 
numerical solutions from the exact solutions (\ref{dnform}) and (\ref{cnform}) 
with parameters found from Newton's method for $c \in (-1,\infty)$ and $m \in [0,m_0)$. Three solution families are shown for $b$ versus $c$ 
for fixed values of $m$. The solid curves represent solutions of the form \eqref{cnform} and the dotted curves are solutions of the form \eqref{dnform}. The black and blue dots demark the points at which the two solution forms connect. Continuing one solution form across these points is impossible because the elliptic modulus $k$ becomes complex-valued.  The red dot shows pitchfork bifurcation (for $m = 0$) 
and fold bifurcation (for $m \neq 0$) when the solution families coalesce.
	
Figure \ref{fig:bifurs}(A) agrees with Figure 2 (middle left panel) in \cite{NLP2} obtained from numerical solutions of the stationary equation (\ref{stationary}). For $m=0$, there exists $c_0 \approx 1.425$ (red dot) at which the pitchfork bifurcation occurs. For $m \neq 0$ in Figures \ref{fig:bifurs}(B-D), the symmetry is broken, the bifurcation point $c_*(m)$ such that $c_*(m) \to c_0$ as $m \to 0$ detaches from the upper branch but remains the connection point for the middle and lower solution families. As $m$ increases, $c_*(m)$ at the bifurcation point increases rapidly and the two solution families move to larger values of $c$.

\begin{figure}[htb!]
	\centering
	\begin{subfigure}[b]{0.5\textwidth}
		\centering
		\includegraphics[width=1\textwidth]{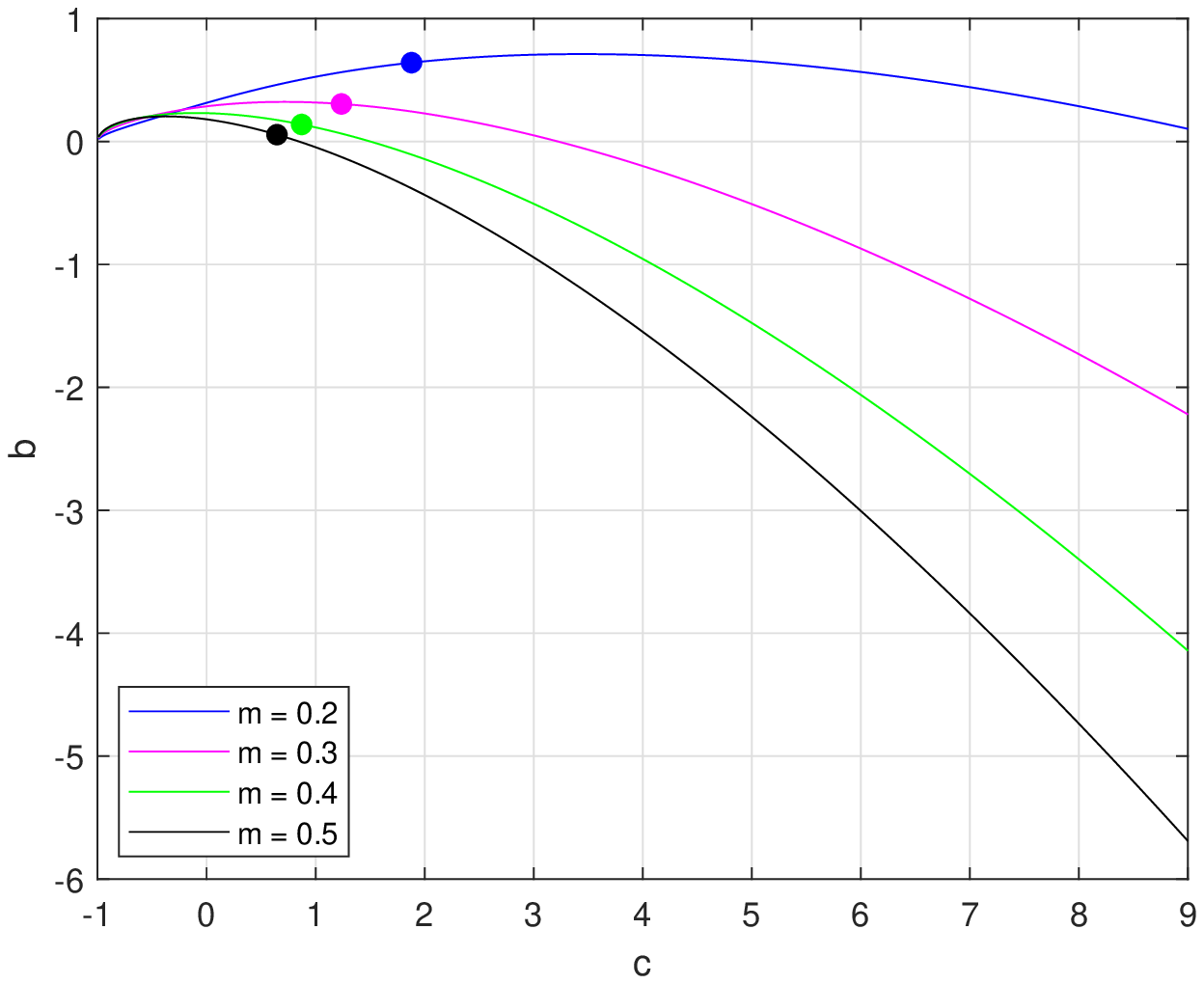}
		\caption{}
		\label{fig:multiple m}
	\end{subfigure}
	\begin{subfigure}[b]{0.5\textwidth}
		\centering
		\includegraphics[width=1\textwidth]{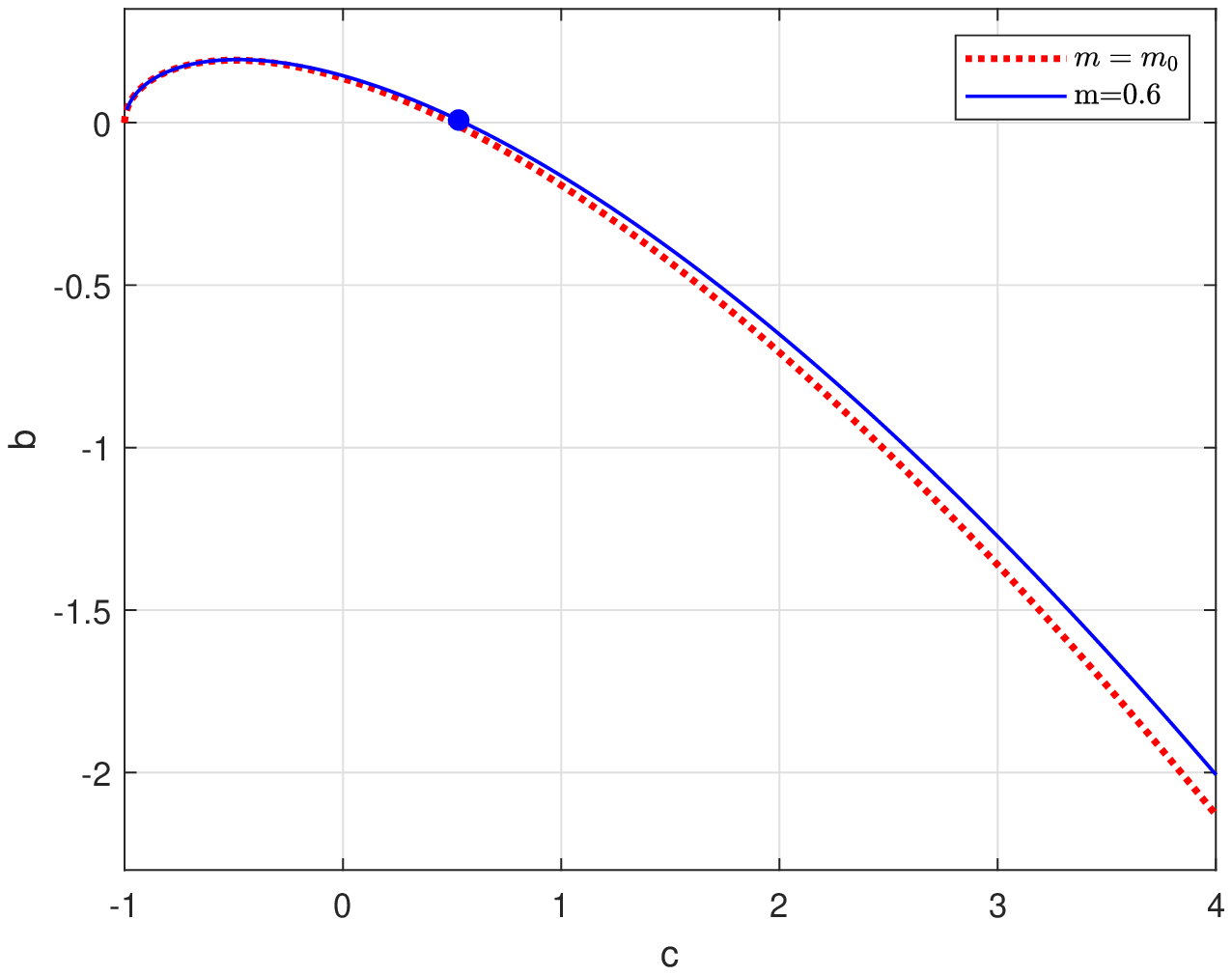}
		\caption{}
		\label{fig:exact m}
	\end{subfigure}
	\caption{(A) The same as Figure \ref{fig:bifurs} but for the upper solution family only for $m = 0.2, 0.3, 0.4, 0.5$. (B) Comparison between the upper solution family for $m = 0.6$ and the exact solution for $m = m_0$.}
	\label{fig:upper}
\end{figure}	

\begin{figure}[htb!]
	\centering
	\begin{subfigure}[b]{0.5\textwidth}
		\centering
		\includegraphics[width=1\textwidth]{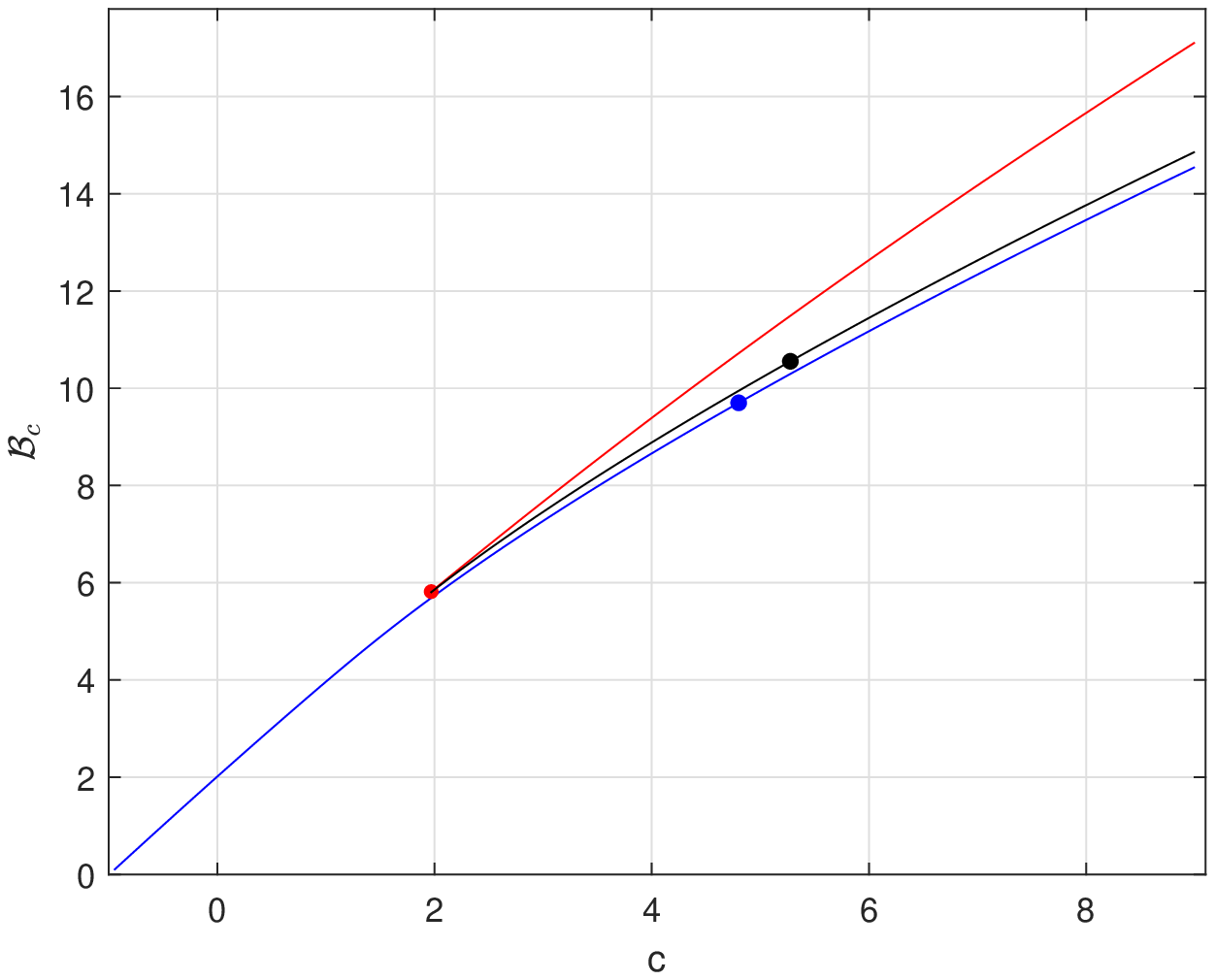}
		\caption{$m=0.01$}
	\end{subfigure}
	\begin{subfigure}[b]{0.5\textwidth}
		\centering
		\includegraphics[width=1\textwidth]{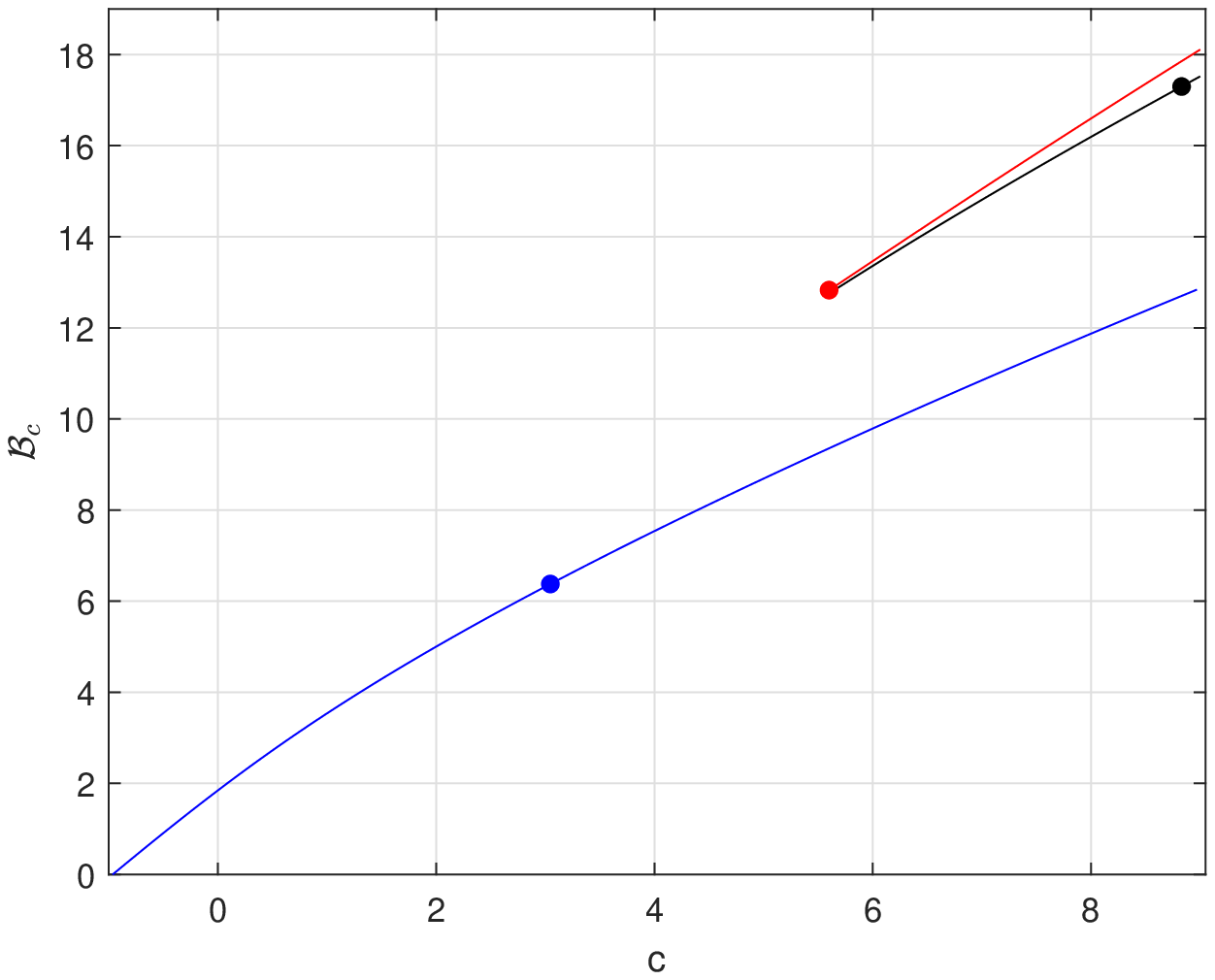}
		\caption{$m = 0.1$}
	\end{subfigure}
	\caption{The value of $\mathcal{B}_c(\chi)$ for the three critical 
		points of the variational problem (\ref{2constraints}) 
		versus $c$ for two values of $m$.}
	\label{fig:minimizers}
\end{figure}
	
Figure \ref{fig:upper}(A) shows only the upper solution family on the $(b,c)$ plane but for larger values of $m$ compared to Fig. \ref{fig:bifurs}. Figure \ref{fig:upper}(B)  compares the numerical result for $m = 0.6$ with 
the analytical result for $m = m_0 \approx 0.6316$ for which the solution family for the constant solutions is given in the parametric form 
\begin{equation}
c = 6 u_1^2 - 1, \quad b = u_1 - 4 u_1^3, \quad u_1 \in (0,\infty). 
\end{equation}
This exact solution for $m = m_0$ follows from either (\ref{dnform}) or (\ref{cnform}) with $u_2 = u_1$.
Although we do not distinguish between the two solution forms (\ref{dnform}) and (\ref{cnform}) in the same solid lines on Figure \ref{fig:upper}, the connection point between the two solutions is shown and it moves to smaller values of $c$ as $m$ increases. 
	
Figure \ref{fig:minimizers} clarifies the meaning of each of the three solution families among the critical points of the variational problem (\ref{2constraints}). It shows the values of $\mathcal{B}_c(\chi)$ defined in \eqref{functionalB} versus $c$ for fixed values of $m = 0.01$ (left) and $m = 0.1$ (right), where $\chi$ is computed from $\psi$ by using $\chi = \psi/\| \psi\|_{L^4}$. To compute $\mathcal{B}_c(\chi)$, we use forward finite difference to approximate $\chi'$ and then complete the quadrature using trapezoidal rule. In both panels, the blue, red and black curves correspond respectively to the upper, middle and lower families of solutions at the bifurcation diagrams of Figure \ref{fig:bifurs}. We observe that the blue curves represent the global minimizers, the black curves  represent the local minimizers, and the red curves represent the saddle points. The numerical search shows that no other solutions 
of the stationary equation (\ref{stationary}) yield critical points
of the variational problem (\ref{2constraints}).

In order to apply the stability criterion (\ref{stability criterion}) for non-degenerate minimizers of the variational problem (\ref{2constraints}), 
we glue individual computations together and represent the solution surface of $b = b(c,m)$ versus $(c,m)$.

\begin{figure}[htb!]
	\centering
	\begin{subfigure}[b]{0.46\textwidth}
		\centering
		\includegraphics[width=1\linewidth]{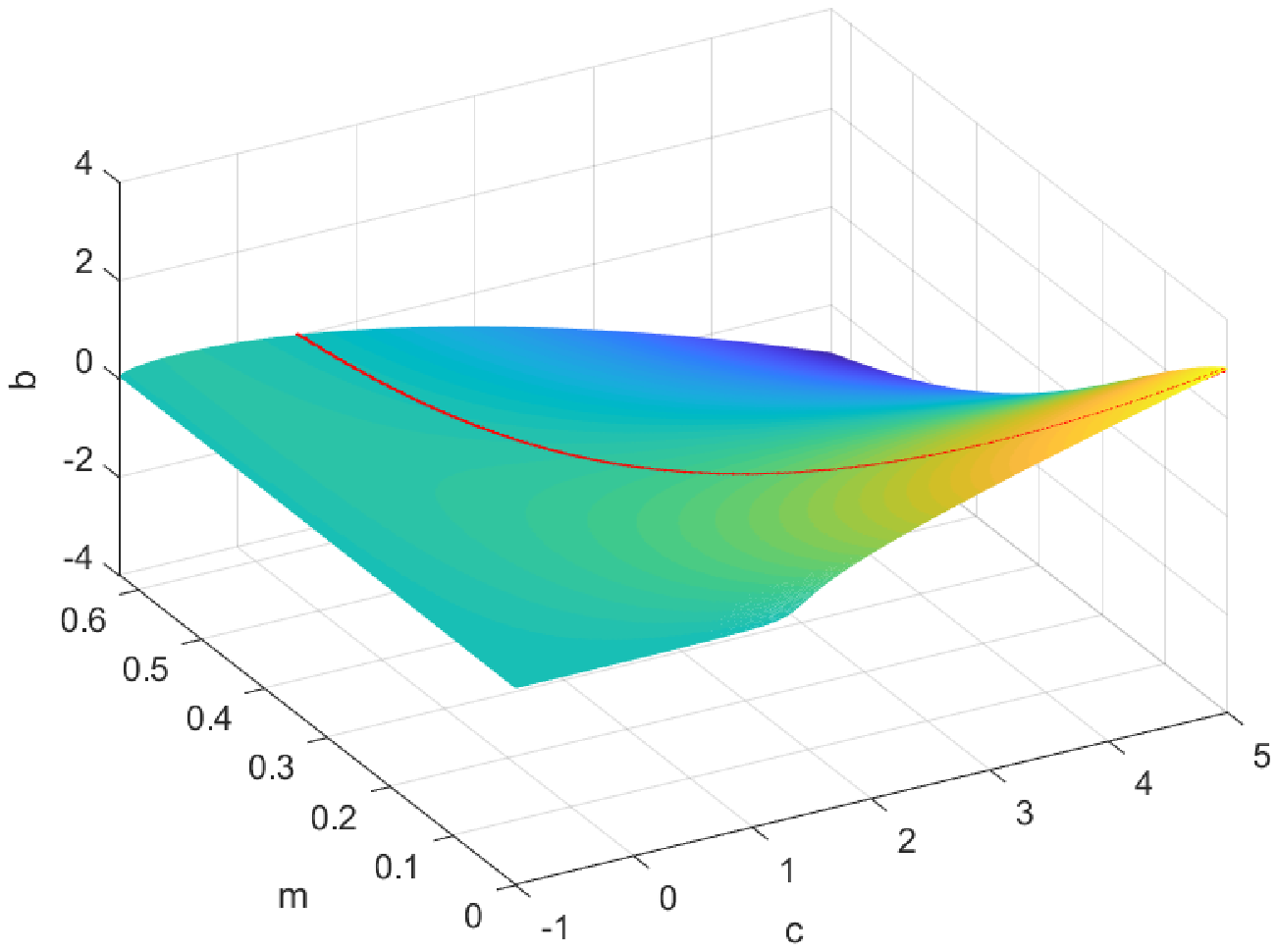}%
		\caption{}
		\label{fig:surface-min}
	\end{subfigure}
	\begin{subfigure}[b]{0.46\textwidth}
		\centering
		\includegraphics[width=1\linewidth]{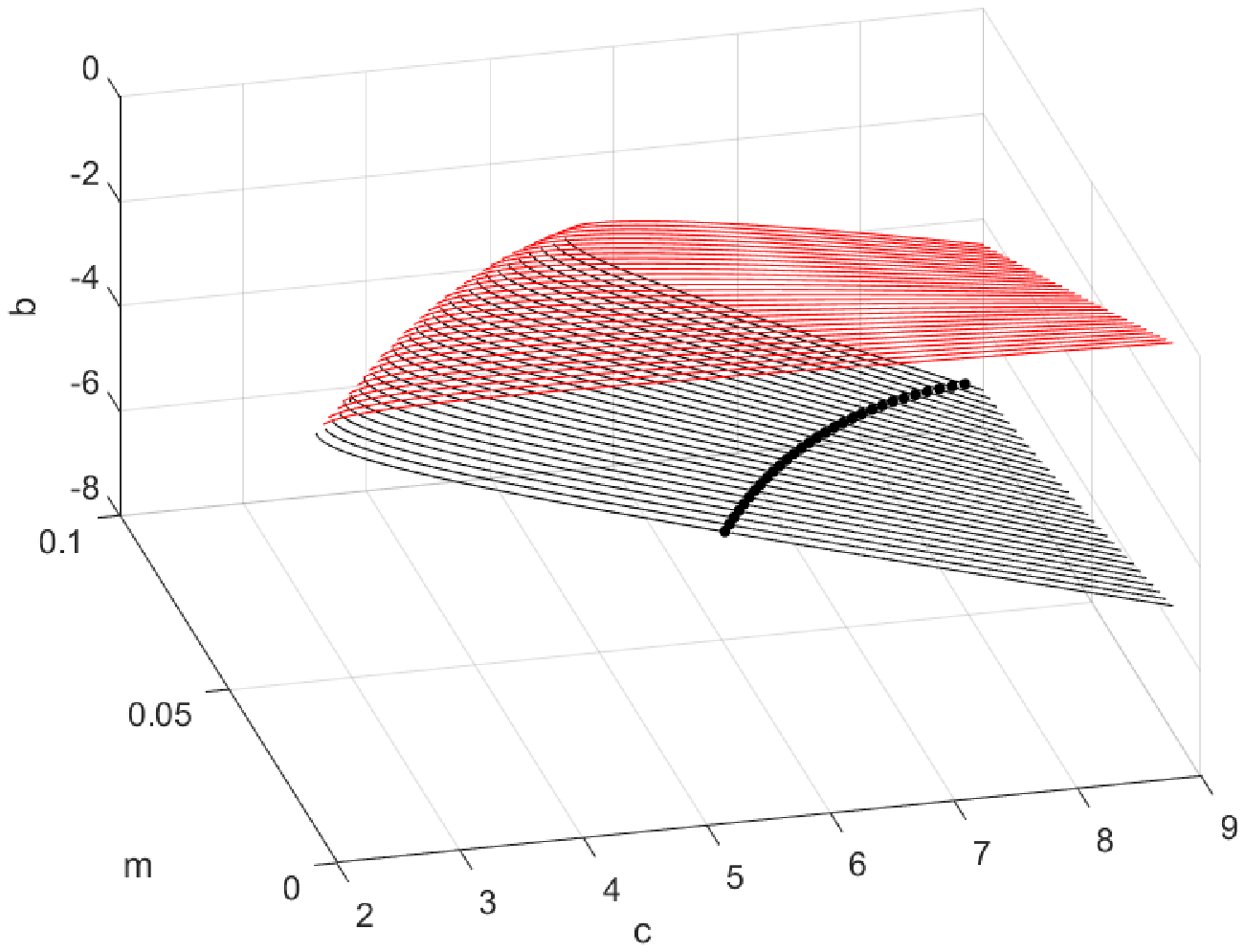}
		\caption{}
		\label{fig:surface-saddle}
	\end{subfigure}
	\caption{(A) The solution surface of $b(c,m)$ for the global minimizers of the variational problem (\ref{2constraints}). (B) Two solution surfaces of $b(c,m)$ for the local minimizers and saddle points of the variational problem (\ref{2constraints}) which connect at the fold bifurcation. }
	\label{fig:surfaceb}
\end{figure} 

Figure \ref{fig:surfaceb}(A) shows the smooth solution surface for the global minimizers of the variational problem (\ref{2constraints}) 
given by the upper solution family on Figure \ref{fig:bifurs} 
for $c \in (-1,\infty)$ and $m \in (0,m_0)$. It suggests non-degeneracy 
of the global minimizers except at the point of the fold bifurcation for $m = 0$ and $c = c_0 \approx 1.425$. The red curve on the solution surface in Figure \ref{fig:surfaceb}  denotes the connection line between the two solution forms \eqref{dnform} and \eqref{cnform}. 

Figure \ref{fig:surfaceb}(B) shows the solution surface $b(m,c)$ for the other two critical points of the variational problem  (\ref{2constraints}). The top (red) part of the surface corresponds to the saddle points and the bottom (black) part of the surface relates to the local minimizers. The numerical result also suggests that the surface is smooth except at the points of the fold bifurcation where the saddle points connect with the local minimizers. The black line denotes the connection line between the two solution forms \eqref{dnform} and \eqref{cnform}.

\subsection{Stability of the global minimizers}

\begin{figure}[htb!]
	\centering
	\includegraphics[width=1\linewidth]{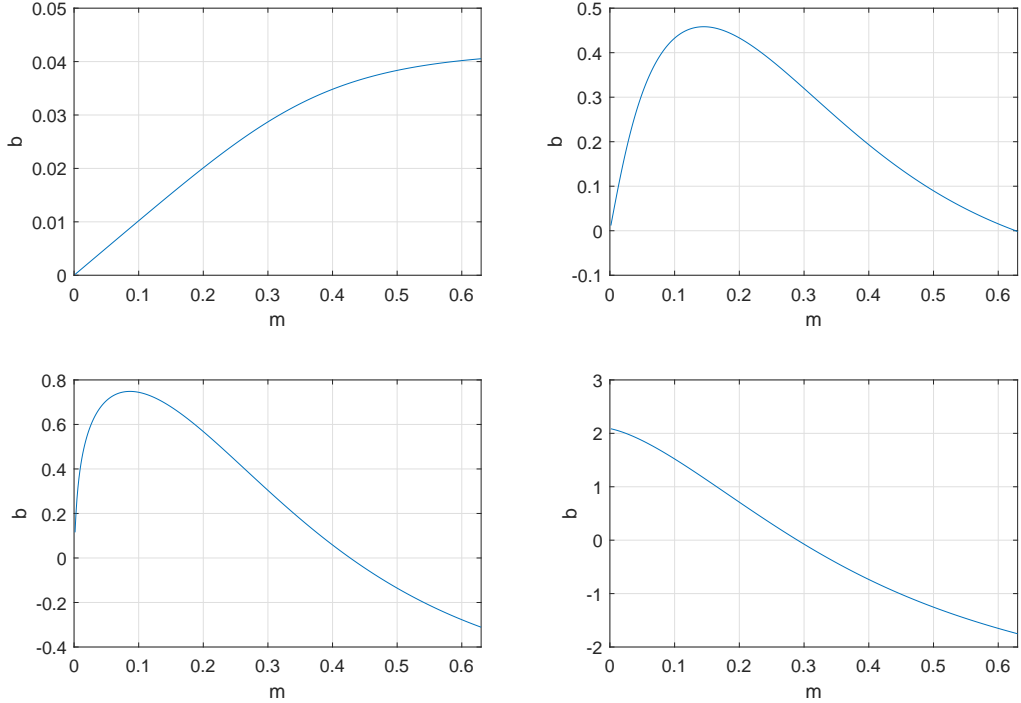}
	\caption{$b$ versus $m$ at the solution surface of Fig. \ref{fig:surfaceb}(A) for various values of $c =-0.99$ (top left), $c = 0.51$ (top right), $c= 1.26$ (bottom left) and $c= 3.53$ (bottom right).}
	\label{fig:snapshotsb}
\end{figure}

It follows from (\ref{Morse}) that the Morse index $n(\mathcal{L})$ and the degeneracy index $z(\mathcal{L})$ depend on the derivative $\partial_m b$.
Figure \ref{fig:snapshotsb} shows $b$ versus $m$ for fixed values of $c$. For $m = 0$,  the derivative $\partial_m b$ changes sign from positive to negative at $c_1 \approx 3.1$. According to (\ref{Morse}), it corresponds to the change in the Morse index $\mathcal{L}$ from 
$n(\mathcal{L}) = 2$ for $c \in (-1,c_1)$ to $n(\mathcal{L}) = 1$ for $c \in (c_1,\infty)$. This agrees with Figure 2 (bottom left panel) in \cite{NLP2}.

For fixed $c \in (-1,\infty)$, it follows from Figure \ref{fig:snapshotsb} 
that there exists $m_1(c)$ for $c \in (-1,c_1)$ such that 
$n(\mathcal{L}) = 2$ for $m \in (0,m_1(c))$ and $n(\mathcal{L}) = 1$ for $m \in (m_1(c),m_0)$. Because $z(\mathcal{L}) = 2$ at $m = m_1(c)$, the non-degeneracy assumption used in the conventional stability theory 
for periodic waves (see \cite{hur,NLP} and references therein) is not satisfied at $m = m_1(c)$. In particular, minimizers of energy $E(u)$ for fixed momentum $F(u)$ and mass $M(u)$ are not smooth with respect to $(c,b)$ at the degeneracy point $m = m_1(c)$. This drawback of the conventional stability theory is not present for the minimizers of the new variational problem (\ref{2constraints}).

\begin{figure}[htb!]
	\centering
	\includegraphics[width=15cm,height=4.5cm]{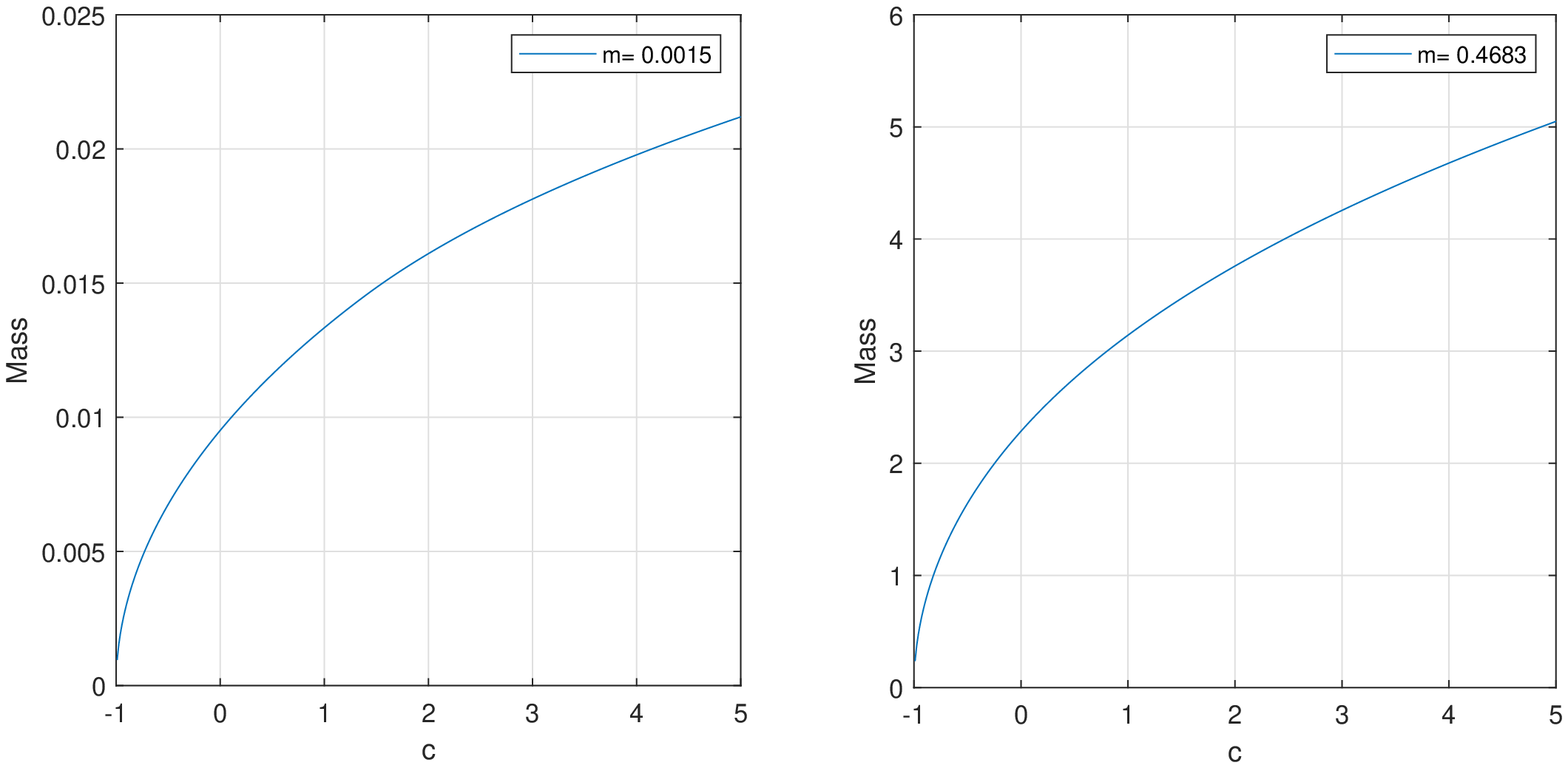}
	\includegraphics[width=15.55cm,height=4.5cm]{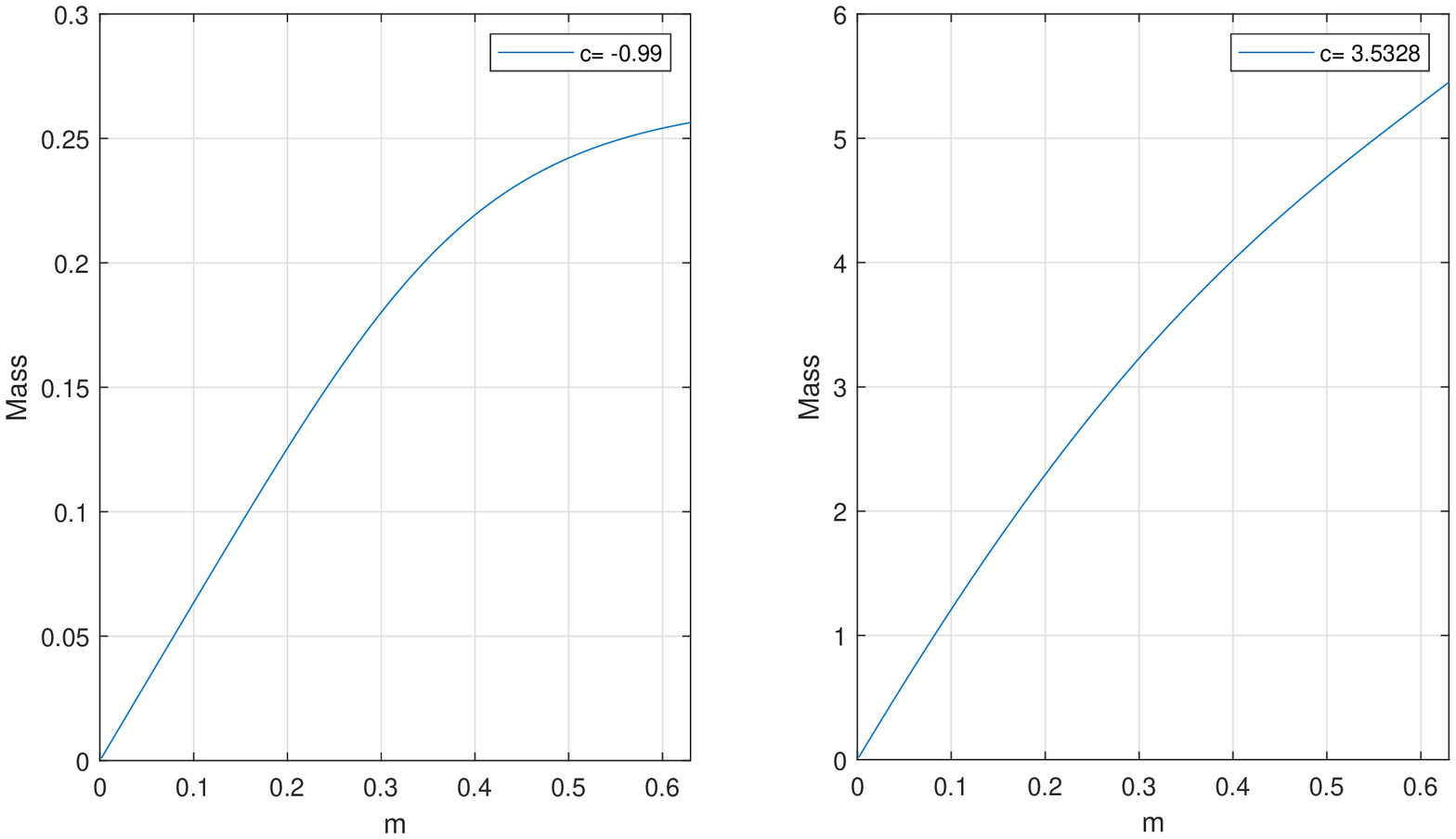}
	\caption{Top: Mass $\mathcal{M}(c,m)$ versus $c$ for various values of $m$. Bottom: Mass $\mathcal{M}(c,m)$ versus $m$ for various values of $c$.}
	\label{fig:snapshotsmass}
\end{figure}

\begin{figure}[htb!]
	\centering
	\includegraphics[width=15cm,height=4.5cm]{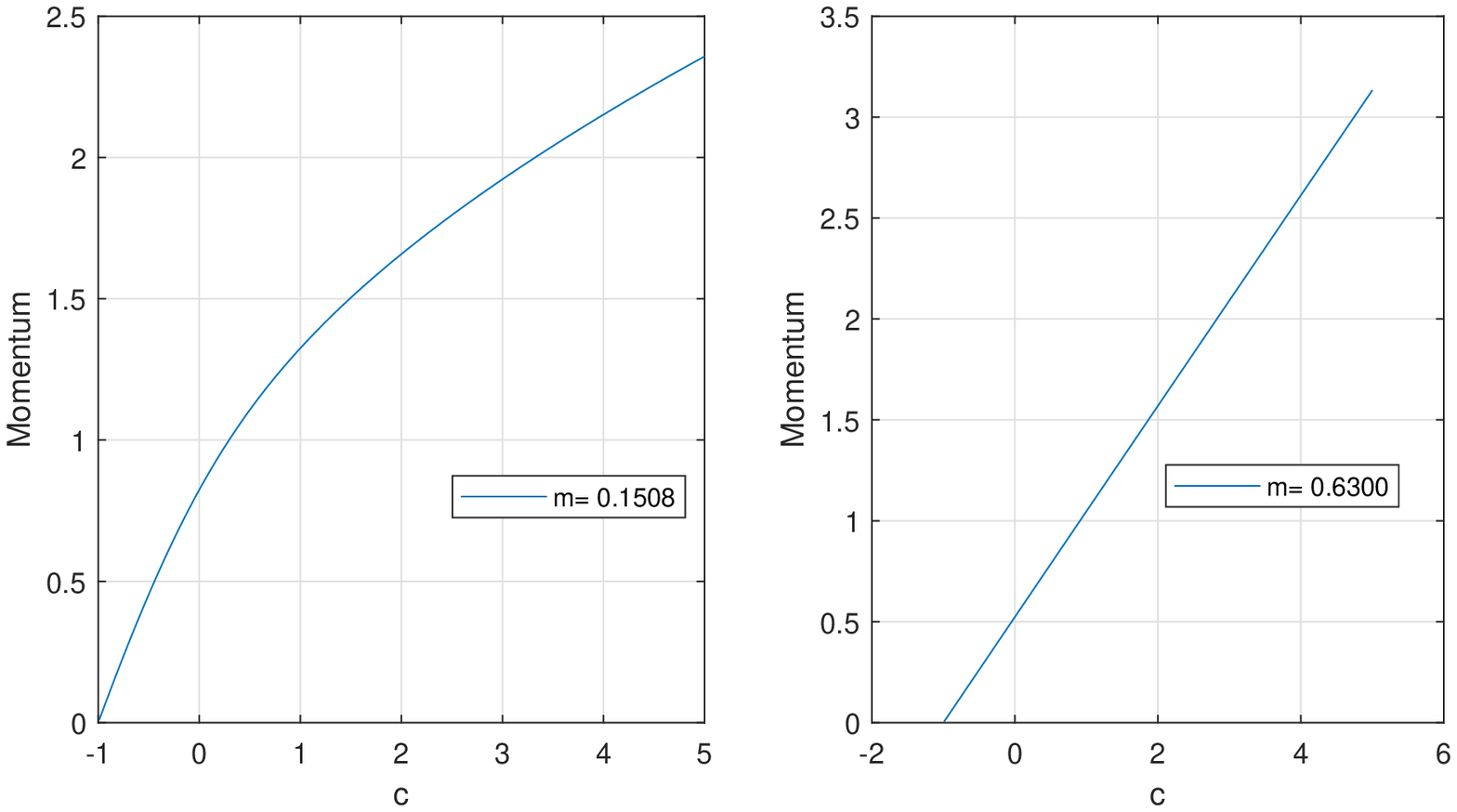}
	\includegraphics[width=14.5cm,height=4.5cm]{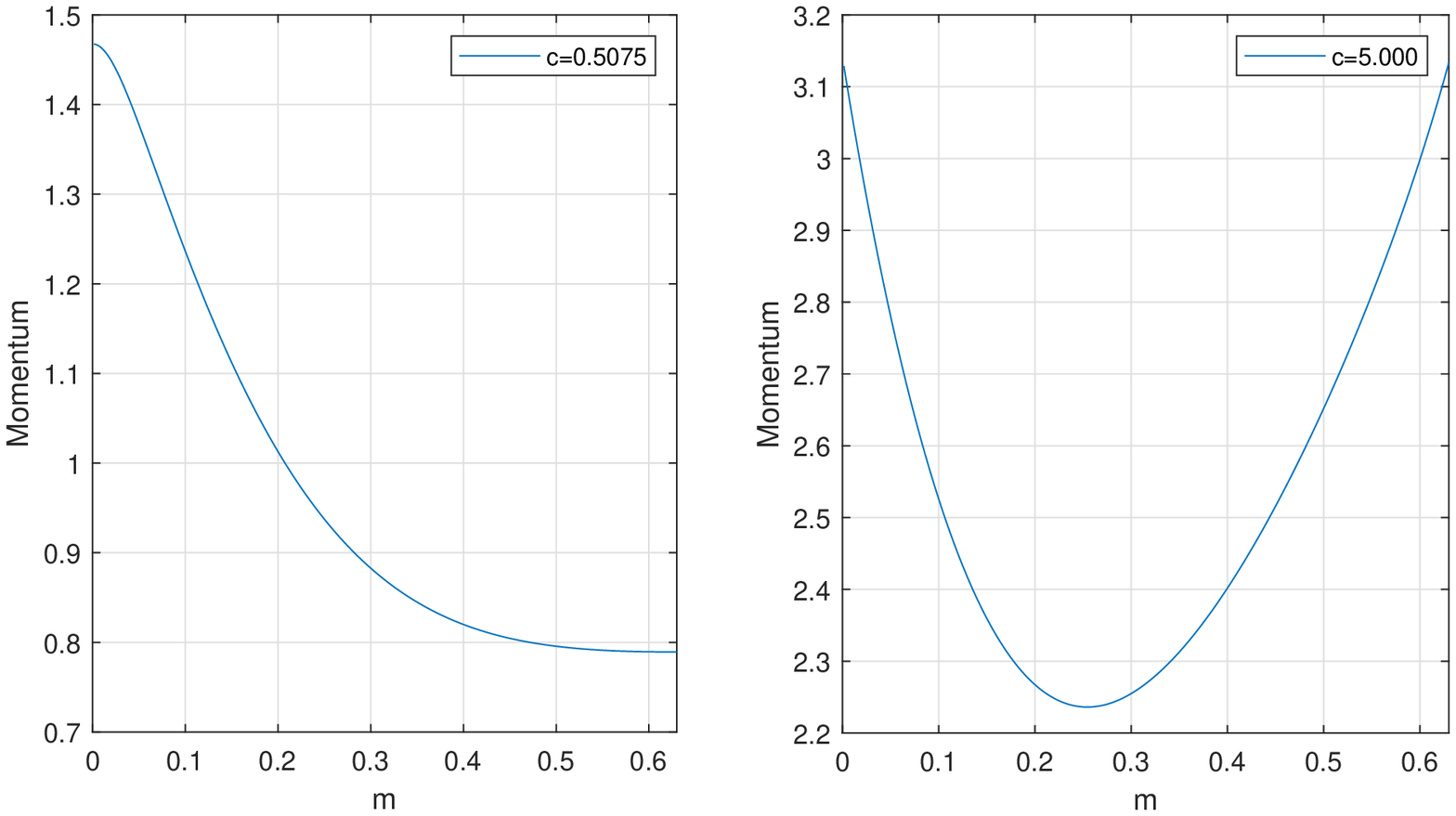}
	\caption{Top: Momentum $\mathcal{F}(c,m)$ versus $c$ for various values of $m$. Bottom:  Momentum $\mathcal{F}(c,m)$ versus $m$ for various values of $c$.}
	\label{fig:snapshotsmomentum}
\end{figure}

Figures \ref{fig:snapshotsmass} and \ref{fig:snapshotsmomentum}  show respectively the mass $\mathcal{M}(c,m)$ and the momentum $\mathcal{F}(c,m)$ versus $c$ at different values of $m$ (top) and versus $m$ at different values of $c$ (bottom). It is clear that the mass $\mathcal{M}(c,m)$ is monotonically increasing in both $c$ and $m$, whereas 
the momentum $\mathcal{F}(c,m)$ is monotonically increasing in $c$ for every $m \in (0,m_0)$ and monotonically decreasing in $m$ for every $c \in (0,c_0)$, where $c_0$ is the same bifurcation value of $c$ for the pitchfork bifurcation at $m = 0$. With these signs of partial derivatives, the stability condition (\ref{stability criterion}) is always satisfied for $c \in (0,c_0)$ and $m \in (0,m_0)$. Note that $\partial_m \mathcal{F}(c,0) = 0$ and $\partial_m \mathcal{M}(c,0) > 0$ so that the stability criterion (\ref{stability criterion}) reduces to $\partial_c \mathcal{F}(c,0) > 0$, monotonicity of the mapping $c \mapsto \mathcal{F}(c,0)$, which was the main stability criterion used in \cite{NLP} and \cite{NLP2}.

\begin{figure}[htb!]
	\centering
	\includegraphics[width=0.6\linewidth]{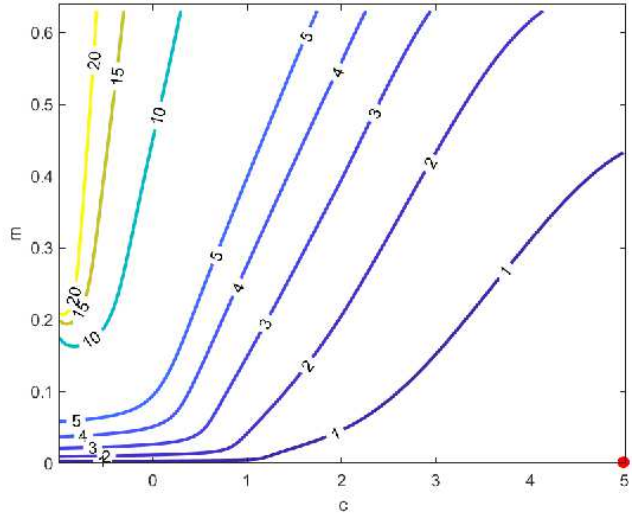}
	\caption{Contour plot of the Jacobian in the stability criterion (\ref{stability criterion}) for the global minimizers.}
	\label{fig:surfacejacobian}
\end{figure}

It follows from Figure \ref{fig:snapshotsmomentum} that for  $c > c_0$, there exists $m_*(c) \in (0,m_0)$ such that 
the momentum $\mathcal{F}(c,m)$ is monotonically decreasing in $m$ for $m \in (0,m_*(c))$ and monotonically increasing in $m$ for $m \in (m_*(c),m_0)$. It is not obvious in the latter case if the stability criterion (\ref{stability criterion}) is satisfied. Figure \ref{fig:surfacejacobian} shows the contour plot for the Jacobian in (\ref{stability criterion}) for all $c \in (-1,5)$ and $m \in (0,m_0)$, which is strictly positive with the minimal value of $0.0145$ attained at the corner point shown by the red dot. Therefore, the stability criterion (\ref{stability criterion}) is satisfied for every non-degenerate minimizer of the variational problem (\ref{2constraints}).

\subsection{Instability of saddle points}

Saddle points of the variational problem \eqref{2constraints} corrrespond to the middle solution family on Figure \ref{fig:bifurs}. The solution surface for the saddle point connects to the solution surface for the local minimizers according to Figure \ref{fig:surfaceb}(B).

\begin{figure}[htb!]
	\centering
	\includegraphics[width=16cm,height=6cm]{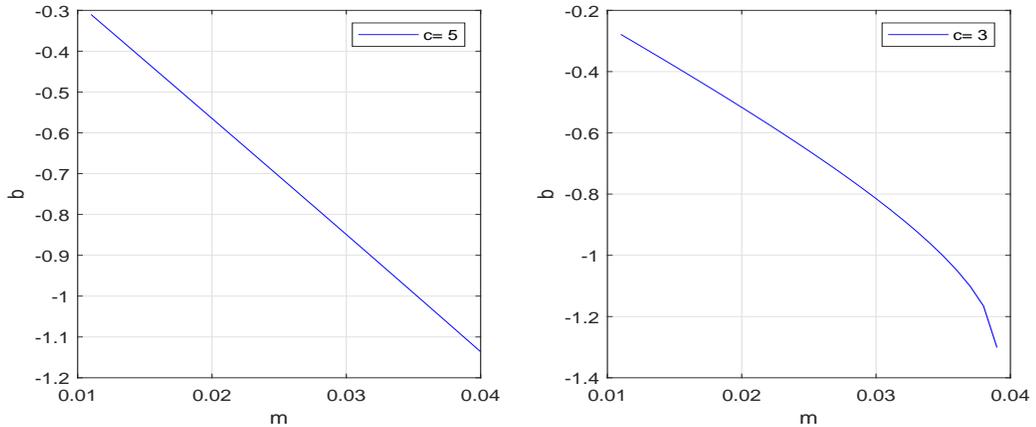}
	\caption{$b$ versus $m$ for the solution surface of Fig. \ref{fig:surfaceb}(B) for $c= 5$ and $c=3$.}
	\label{fig:snaps-bmiddle}
\end{figure}

The instability criterion for the saddle points (\ref{instability criterion}) was derived under the assumption of $\partial_m b < 0$. Figure \ref{fig:snaps-bmiddle} shows $b$ versus $m$ for two values of $c$, which suggests that $\partial_m b <0$ is satisfied for all saddle points of the variational problem (\ref{2constraints}).

We have found numerically that the mass $\mathcal{M}(c,m)$ of the saddle points is monotonically increasing in both $m$ and $c$ similar to Figure \ref{fig:snapshotsmass} for the global minimizers. We also found 
that the momentum $\mathcal{F}(c,m)$ is monotonically increasing in $c$ 
but there exist $c_1$ and $c_2$ satisfying $3 < c_1 < c_2 < 9$ such that $\mathcal{F}(c,m)$ is monotonically decreasing in $m$ for $c < c_1$ and monotonically increasing in $m$ for $c > c_2$ similar to Figure \ref{fig:snapshotsmomentum}. Although the sign of the Jacobian in \eqref{instability criterion} is not obvious in the latter case, we have computed it numerically and confirmed that the Jacobian is strictly positive for the entire solution surface. Thus, the saddle points of the variational problem (\ref{2constraints}) are unstable in the time evolution of the mKdV equation (\ref{mKdv}) according to the instability criterion \eqref{instability criterion}.

\subsection{Stability of local minimizers}

Local minimizers of the variational problem \eqref{2constraints} corrrespond to the lower solution family on Figure \ref{fig:bifurs}. We have checked numerically 
that the plots of $b$ versus $m$, $\mathcal{M}(c,m)$ and $\mathcal{F}(c,m)$ versus both $c$ and $m$ are qualitatively similar 
to Figures \ref{fig:snapshotsb}, \ref{fig:snapshotsmass}, and \ref{fig:snapshotsmomentum}. This is not surprising since 
the lower and upper solution families are equivalent to each other for $m = 0$. 

We have also detected numerically that the Jacobian in the stability criterion 
(\ref{stability criterion}) remains positive for the entire solution surface. 
Thus, the local minimizers of the variational problem (\ref{2constraints})  are stable in the time evolution of the mKdV equation (\ref{mKdv}) according to the stability criterion \eqref{stability criterion}.

\section{Conclusion}

The new variational characterization of periodic waves as non-degenerate minimizers of the variational problem (\ref{2constraints}) has several advantages compared to the previous variational theory, where the energy $E(u)$ is minimized for fixed momentum $F(u)$ and mass $M(u)$. First, 
the stability criterion is independent of whether the Morse index $n(\mathcal{L})$ is one or two and whether the linear operator $\mathcal{L}$ is degenerate with $z(\mathcal{L}) = 2$. Second, with the exception of the 
pitchfork bifurcation point $(c,m) = (c_0,0)$, minimizers of the 
variational problem (\ref{2constraints}) are always non-degenerate.

The new variational characterization has also advantages compared to other (partial) characterizations of periodic waves in the mKdV equation such as 
minimization of energy $E(u)$ for fixed momentum $F(u)$ in \cite{stefanov} 
or minimization of $B_c(u)$ for fixed $L^4$ norm in the space of even functions 
also considered in \cite{NLP2}. The former minimization only allows to identify a subset of stable periodic waves as it only applies to the periodic waves with $n(\mathcal{L}) = 1$. The latter minimization requires to proceed with an additional Galilean transformation in order to identify the stability criterion for the periodic waves and leads to computational formulas which are not related to the dependence of mass $M(u)$ or momentum $F(u)$ on $c$.

Although our computations are only based on numerical approximations, 
the numerical results are rather accurate since we use the exact analytical 
representations of the periodic wave solutions. We have confirmed that 
the periodic waves that correspond to the global and local minimizers of the variational problem (\ref{2constraints}) are stable in the time evolution 
of the mKdV equation (\ref{mKdv}), whereas the periodic waves for the saddle points are unstable.  

The main direction to be addressed in further work is to prove analytically that 
minimizers of the variational problem (\ref{2constraints}) are non-degenerate 
in the entire existence interval with the exception of the pitchfork bifurcation 
point at $(c,m) = (c_0,0)$. Extensions of these numerical results to the modified Benjamin--Ono equation or the fractional mKdV equation 
are also of the highest priority. Finally, one can apply the same new variational problem to the generalized fractional KdV equations with powers different from the quadratic and cubic powers considered in \cite{NLP} and \cite{NLP2} respectively.

\vspace{0.25cm}

{\bf Acknowledgments.} U. Le acknowledges the support of McMaster graduate scholarship. D.E. Pelinovsky acknowledges the support of the NSERC Discovery grant.

\end{document}